\documentclass[aps, twocolumn]{revtex4-2}
\usepackage{epsfig}
\usepackage{graphicx}
\usepackage{amsmath,amssymb,amsfonts}
\usepackage{array}
\usepackage{url}
\usepackage[usenames,dvipsnames]{color}
\usepackage[colorlinks=true,linkcolor=magenta, citecolor = blue]{hyperref}
\usepackage{multirow}
\usepackage{float}
\usepackage{lineno}
\usepackage{xspace}
\usepackage{ulem}
\usepackage{lipsum} 
\usepackage{bigints}
\begin{document}
\title{Dissociation-driven quarkonium spin alignment in Pb--Pb collisions \\ at $\sqrt{s_{\rm NN}} = 5.02$ TeV}
\author{Bhagyarathi Sahoo}
\email{Bhagyarathi.Sahoo@cern.ch}
\author{Captain R. Singh}
\email{captainriturajsingh@gmail.com}
\author{Raghunath Sahoo}
\email{Raghunath.Sahoo@cern.ch (Corresponding author)}
\affiliation{Department of Physics, Indian Institute of Technology Indore, Simrol, Indore 453552, India}

\begin{abstract}
The observation of spin alignment of quarkonia in ultra-relativistic heavy-ion collisions provides deep insight into the 
possible formation of the quark-gluon plasma (QGP). The present study investigates the spin alignment of quarkonia 
induced by dissociation mechanisms arising from medium effects imposed on quarkonia. We implement an effective 
Hamiltonian with a medium-modified color-singlet potential to incorporate the coupling of quarkonium spin with medium 
vorticity. This coupling gives rise to spin-dependent dissociation, which we identify as a plausible mechanism 
contributing to quarkonium spin alignment. Within the ambit of second-order relativistic viscous hydrodynamics, we 
calculate the spin-dependent decay widths of charmonium ($J/\psi$, $\psi$(2S)) and bottomonium ($\Upsilon$(1S), 
$\Upsilon$(2S)) in a rotating thermal medium, including collisional damping and gluonic dissociation effects. We evaluate 
the observable $\rho_{00}$ for Pb--Pb collisions at $\sqrt{s_{\rm NN}} = 5.02$ TeV as a function of transverse momentum of 
the quarkonia, charged particle multiplicity, and medium rotation. The results demonstrate that medium vorticity modifies the quarkonia net 
decay width and, as a consequence, quarkonia spin alignment gets modified. These findings suggest new directions for 
understanding spin transport and the microscopic dynamics of vortical QGP.

\end{abstract}
\date{\today}
\maketitle

\section{Introduction}

The hunt for the deconfined phase of strongly interacting matter known as quark-gluon plasma (QGP) in ultra-relativistic heavy-ion collisions was at its peak, but then something unexpected happened in the grand caves under the Jura mountains. 
The evidence for QGP formation in heavy-ion collisions has been sought through observables such as collective flow, 
strangeness enhancement, quarkonia suppression, etc.~\cite{Matsui:1986dk, ALICE:2012jsl, CMS:2012bms, ALICE:2013xmt, ALICE:2016ccg}. Similar to those, a substantial influence of the medium on 
observables, like strangeness enhancement and collective flow at high multiplicities in $p-p$ collisions, was observed at 
the LHC energies~\cite{ALICE:2016fzo, CMS:2016fnw, ALICE:2024vzv}. These findings suggest the possible emergence of QGP-like phenomena in small systems, thereby opening a 
new research domain to investigate the existence of QGP in such small collision systems. These findings also pose a 
challenge to established methodologies, particularly for observables that use $p-p$ collisions as a baseline, like 
quarkonia suppression. However, quarkonia, being the bound states of the heavy quark and antiquark pair, have masses significantly larger than 
the QCD scale, $\Lambda_{\rm QCD}$. The large heavy quark masses originate mainly from their Yukawa couplings to the 
Higgs field through electroweak symmetry breaking, in contrast to light quarks whose effective masses are largely 
generated by non-perturbative QCD dynamics and spontaneous chiral symmetry breaking. Therefore, quarkonia, with their 
unique properties, such as large binding energies, small spatial radii, and distinct sequential melting temperatures, 
make them a better probe for investigation of deconfined QCD matter ~\cite{Matsui:1986dk}. Since quarkonium states are produced predominantly 
during the early hard scatterings, the vorticity produced in the earliest times of collisions is expected to affect heavy quarks   (antiquarks) 
more strongly than light flavors. Through spin-orbit coupling, such early-time vorticity can induce polarization 
in quarkonium states, which would access the full space–time evolution of the system, and may provide a direct connection 
between the early dynamics and the final states.\\

Recent experimental measurements of hyperon spin polarization and vector meson spin alignment have initiated a more 
focused study on hadron polarization in ultra-relativistic heavy-ion collisions~\cite{STAR:2017ckg, STAR:2020xbm, STAR:2022fan, STAR:2023nvo, ALICE:2019aid, ALICE:2025cdf}. Substantial progress has been made to comprehend global, local, and transverse spin polarization of various 
hadrons from both theoretical and 
experimental standpoint~\cite{Liang:2004ph, Liang:2004xn, Karpenko:2016jyx, Alzhrani:2022dpi, Teryaev:2017wlm, Sheng:2019kmk, Sheng:2022wsy, Sahoo:2024yud, Sahoo:2024egx, Sahoo:2025bkx, Sahoo:2023oid, LHCb:2025rxf, STAR:2025jwc}. In this context, the first measurement of $J/\psi$ spin alignment in Pb--Pb collisions at $\sqrt{s_{\rm NN}}$ = 
5.02 TeV has revealed considerable polarization signals in multiple reference frames, including the helicity, Collins-Soper, and event-plane frames ~\cite{ALICE:2020iev, ALICE:2022dyy}. These observations have stimulated considerable theoretical interest, motivating 
efforts to make predictions consistent with experimental results and to identify the possible underlying mechanisms of 
$J/\psi$ polarization. Among the proposed sources, the vorticity field generated in peripheral collisions is of 
particular importance. Arising from the large initial orbital angular momentum (OAM) of the colliding nuclei, this 
vorticity can polarize quarks through spin-orbit coupling, with the resulting polarization subsequently transferred to 
the produced hadrons ~\cite{Liang:2004ph, Liang:2004xn}.\\

The spin alignment of quarkonia is characterized through the elements of a $3 \times 3$ Hermitian spin density matrix, 
$\rho_{m, m'}$, where $m$ and $m'$ denote the spin projections along a chosen quantization axis. The diagonal elements 
$\rho_{11}$, $\rho_{00}$, and $\rho_{-1-1}$ represent the probabilities of finding the vector meson in spin states $+1$, 
$0$, and $-1$, respectively. Among these, $\rho_{00}$ is of particular importance, as it can be directly measured in the 
experiments through the angular distribution of the decay products. For quarkonia, $\rho_{00}$ is determined from the 
angular distribution of decay leptons in the quarkonium rest frame, which is given ~\cite{Sahoo:2023oid};

\begin{equation}
\frac{dN}{d\Omega} \propto 1 + \lambda_{\theta} \cos^{2}\theta 
+ \lambda_{\phi} \sin^{2}\theta \cos 2\phi 
+ \lambda_{\theta\phi} \sin 2\theta \cos \phi 
\end{equation}

where $\theta$ and $\phi$ are the polar and azimuthal angles of the decay products, while $\lambda_{\theta}$, 
$\lambda_{\phi}$ and $\lambda_{\theta\phi}$ are the polarization parameters that quantify the anisotropy of the decay 
distribution. These parameters are directly related to the elements of the spin density matrix. The parameter 
$\rho_{00}$, which quantifies the probability of the vector meson being in a spin-zero projection along the quantization 
axis, is connected to the polar anisotropy parameter $\lambda_{\theta}$ by the relation,

\begin{equation}
\rho_{00} = \frac{1 - \lambda_{\theta}}{3 + \lambda_{\theta}}
\end{equation}

In the absence of any spin alignment, the three spin substates ($m_j = +1, 0, -1$) are equally probable, giving $\rho_{00} 
= 1/3$. The deviation $\rho_{00} - 1/3$ thus quantifies the degree of alignment of the vector meson. A value $\rho_{00} < 
1/3$ indicates a transverse alignment, whereas $\rho_{00} > 1/3$ corresponds to a longitudinal alignment with respect to 
the chosen quantization axis. A precise determination of $\rho_{00}$ and its associated polarization parameters in 
different reference frames, such as the helicity, Collins–Soper, and event-plane frames, provides key insights into the 
spin dynamics of quarkonia and the influence of the surrounding medium in relativistic heavy-ion collisions. However, the recent experimental measurement of spin alignment of  $J/\psi$ in Pb--Pb collisions at $\sqrt{s_{\rm NN}}$ = 5.02 TeV is not consistent across different polarization frame of reference ~\cite{ALICE:2020iev, ALICE:2022dyy}. The spin alignment observable $\rho_{00} <$ 1/3 for helicity and event plane frame, while it is observed to be $\rho_{00} >$ 1/3 Collins-Soper frame. \\

In this study, we investigate the spin alignment of charmonium and bottomonium states, specifically $J/\psi$, $\psi$(2S), 
$\Upsilon$(1S), and $\Upsilon$(2S) as a function of transverse momentum ($p_{\rm T}$) and charged particle multiplicity in Pb--Pb 
collisions at $\sqrt{s_{\rm NN}} = 5.02~\text{TeV}$. 
This study examines the spin alignment of quarkonia along the vorticity or total angular momentum
direction, which is perpendicular to the reaction plane~\cite{Becattini:2007sr}. In the absence of
fluctuations, the event plane serves as an estimate of the reaction plane~\cite{Bhalerao:2020ulk}.
Therefore, the present calculation is more closely aligned with the event-plane frame. Our study
focuses  on the spin-dependent dissociation of quarkonia as
a possible mechanism for their spin alignment in a vortical QGP medium. Here, we have obtained the spin alignment of 
quarkonium states considering collisional damping and the gluonic dissociation mechanisms. The
combined effect of these
in-medium mechanisms on the spin alignment observable $\rho_{00}$ is systematically explored.
Further, the space-time evolution of the QGP is modeled using a second-order relativistic viscous
hydrodynamic framework, which provides the
temperature cooling profile in the expanding medium. The effect of medium vorticity, characterized
through the conserved
circulation $C$, is further examined to assess its role in generating spin alignment for different
quarkonium states
within the deconfined phase of thermally equilibrated hot and vortical QCD matter.

The paper is organized as follows: Section~\ref{formulation} presents the detailed theoretical formulation used in this 
study. Section~\ref{res} discusses the results and their physical implications. Finally, Section~\ref{sum} summarizes our 
findings and provides an outlook for future research.

\section{Formulation}
\label{formulation}
The evolution of quarkonium states within the QGP medium is described using the time-independent
Schr\"{o}dinger equation (SE) and followed by dissociation mechanisms like collisional damping and gluonic dissociation. 
Collisional damping and gluon-induced dissociation are among the extensively studied processes governing quarkonium 
suppression in the QGP medium~\cite{ Singh:2015eta, Ganesh:2016kug, Singh:2018wdt, Hatwar:2020esf, Singh:2021evv, 
Singh:2025xrd}. These mechanisms induce spin-dependent modifications in their dissociation widths, thereby influencing 
the spin density matrix element, $\rho_{00}$. To quantify this effect, we calculate the spin-dependent decay widths 
arising from both collisional damping and gluonic dissociation by incorporating the coupling between the spin of the 
particle and the vorticity of the medium within an effective Hamiltonian framework that includes a medium-modified 
potential. Given the pivotal role of temperature in modifying the quarkonia characteristics within the QGP medium, it is 
imperative first to examine the thermodynamic evolution of deconfined QCD matter.

\subsection{Thermal Evolution of the Medium}

The possible formation of QGP in ultra-relativistic heavy-ion collisions undergoes rapid expansion in both space and time 
domains due to strong internal pressure gradients. The space–time evolution of such a deconfined medium is proposed to be 
governed by relativistic hydrodynamics. In the present study, we employ the second-order relativistic viscous 
hydrodynamics framework to describe the dynamical evolution of the QGP. To solve the hydrodynamic equations, one must 
specify appropriate initial conditions and an equation of state (EoS) that connects with the thermodynamic quantities of 
the system. In the absence of a first-principles determination of the initial temperature ($T_0$) and thermalization time 
($\tau_0$) of QGP, we estimate $T_0$ using a phenomenological relation~\cite{Hwa:1984fj} that connects it to the 
experimentally measured charged particle multiplicity:

\begin{equation}
 T_{0} = \left[\frac{90}{g_{k} 4\pi^{2}} C^{\prime} \frac{1}{A_{T} \tau_{0}} \frac{dN_{\text{ch}}}{dy} \right]^{1/3},
 \label{t0}
 \end{equation}

here the constant is $C^{\prime} = \frac{2\pi^{4}}{45\zeta(3)} \approx 3.6$, and $\frac{dN_{\text{ch}}}{dy} \simeq 
\frac{dN_{\text{ch}}}{d\eta}$, under the massless limit. Equation~\ref{t0} is derived considering the isentropic 
expansion of the fireball. In a QGP medium with comparatively small shear viscosity, the uncertainties introduced by this assumption are 
negligible compared to those arising from other parameters such as the transverse overlap area ($A_T$), thermalization 
time, and statistical degeneracy ($g_k$) of the deconfined phase.
The transverse overlap area, $A_T = \pi R_T^2$, is estimated using the MC-Glauber mode for Pb--Pb collision at 
$\sqrt{s_{\rm NN}} = 5.02$ TeV, where $R_T$ is the transverse radius of the fireball. 
It is important to emphasize that the initial thermalization time $\tau_0$ plays a decisive role in determining the 
initial temperature $T_0$. In the present work, we adopt $\tau_0 = 0.2~\text{fm}/c$ for Pb--Pb collisions at 
$\sqrt{s_{\text{NN}}} = 5.02~\text{TeV}$, the chosen value is consistent with those used in modern viscous hydrodynamic 
simulations. Corresponding to this choice of $\tau_0$, the values of $T_0$ are obtained as a function of the measured 
charged particle multiplicity. The charged particle multiplicity classes for Pb+Pb collisions at $\sqrt{s_{\rm NN}} = 5.02$ TeV is taken from ALICE experimental data~\cite{ALICE:2015juo} and reported in Table~\ref{Table:multiplicity}.\\ 

\begin{table}[htbp]
\scalebox{1.1}
{
\renewcommand{\arraystretch}{1.5}
\begin{tabular}{c|c|c|c|c|c|}
\hline
\multicolumn{2}{|c|} {{ Multiplicity Class}} &  Centrality Class &    $ \langle \frac{dN_{\text{\rm ch}}}{d \eta} \rangle $\\
\hline
\multicolumn{2}{|c|}{ Mult. Class - 1 } &  (70-80)\% & 44.9 $\pm$  3.4\\
\hline 

\multicolumn{2}{|c|}{ Mult. Class - 2} & (60-70)\% & 96.3 $\pm$  5.8 \\
\hline 

\multicolumn{2}{|c|}{ Mult. Class - 3 } &  (50-60)\% & 183 $\pm$ 8\\
\hline 

\multicolumn{2}{|c|}{ Mult. Class - 4} & (40-50)\% & 318 $\pm$  12 \\
\hline 

\multicolumn{2}{|c|}{ Mult. Class - 5 } &  (30-40)\% & 512 $\pm$  15\\
\hline 

\multicolumn{2}{|c|}{ Mult. Class - 6} & (20-30)\% & 786 $\pm$ 20 \\
\hline 

\multicolumn{2}{|c|}{ Mult. Class - 7 } &  (10-20)\% & 1180 $\pm$  31\\
\hline 

\multicolumn{2}{|c|}{ Mult. Class - 8} & (7.5-10)\% & 1505 $\pm$  44 \\
\hline 

\multicolumn{2}{|c|}{ Mult. Class - 9 } &  (5.0-7.5)\% & 1666 $\pm$  48\\
\hline 

\multicolumn{2}{|c|}{ Mult. Class - 10} & (2.5-5.0)\% & 1850 $\pm$  55 \\
\hline 

\multicolumn{2}{|c|}{ Mult. Class - 11} & (0-2.5)\% & 2035  $\pm$  52 \\
\hline 

\end{tabular}
}
\caption{The $\langle\frac{dN_{\text{\rm ch}}}{d\eta}\rangle$ values
measured for Pb+Pb collisions at $\sqrt{s_{\rm NN}} = 5.02$ TeV in $|\eta| < $ 0.5 for 11 centrality classes are taken from ALICE experimental data~\cite{ALICE:2015juo}.}
\label{Table:multiplicity}
\end{table}

Now, with the initial conditions in hand, we employ 1+1D with boost invariance second-order viscous hydrodynamics to 
determine the temperature cooling profile in the medium~\cite{Muronga:2003ta}. The second-order cooling law is derived 
from kinetic theory using Grad's 14-moment approximation method within the framework of  the M\"{u}ller-Israel-Stewart 
equation~\cite{gam,mis1,mis2,mis3}, given as;

\begin{equation}
\frac{dT}{d\tau} = -\frac{T}{3\tau} + \frac{T^{-3}\phi}{12a\tau}
\label{so1}
\end{equation}
and 
\begin{equation}
\frac{d\phi}{d\tau} = -\frac{2aT\phi}{3b} - \frac{1}{2}\phi\left[\frac{1}{\tau} -\frac{5}{T}\frac{dT}{d\tau}\right] + \frac{8aT^{4}}{9\tau}
\label{so2}
\end{equation}

The variable $\phi$ in Eq.~(\ref{so2}) is called the shear viscous term, which quantifies the time-dependent change in the shear viscosity ($\eta$). In essence, $\phi$ 
reflects the characteristics of the medium created under extreme conditions during ultra-relativistic collisions. For the 
first-order solution of Eq.~(\ref{so1}), $\phi$ is defined as $\phi = 4\eta/3\tau$. The constants $a$ and $b$ are defined 
as;

\begin{equation}
a = \frac{\pi^{2}}{90} \left[ 16 + \frac{21}{2}N_{f} \right]
\label{a}
\end{equation}
and
\begin{equation}
b = (1 + 1.70 N_{f})\frac{0.342}{(1 + N_{f}/6)\alpha_{s}^{2}\ln(\alpha_{s}^{-1})}
\label{b}
\end{equation}
here $N_{f} = 3$, is the number of flavors and $\alpha_s$ is the strong coupling constant.
The first-order viscous correction term, $\phi = 4\eta / (3\tau)$, is used to set the initial condition for 
Eq.~(\ref{so2}). By taking the KSS bound of the shear viscosity to entropy density ratio, $\eta/s = 1/(4\pi)$, the initial 
value of the viscous term at $\tau_0$ can be expressed as $\phi_{0} = (1/3\pi)(s_{0}/\tau_{0})$. The initial entropy 
density ($s_{0}$) is estimated using the quasi-particle model (QPM) equation of state at zero baryon chemical potential~\cite{Singh:2018wdt}.

\subsection{Quarkonia in-medium}
Quarkonia traversing through the medium are supposed to be affected by the surrounding partonic environments. The influence of the medium on quarkonia depends on how fast quarkonia is moving in the medium and the relative velocity between the medium and quarkonia. Apart from this, quarkonia, being heavy, do not share the same temperature as the medium. Therefore, to account for the precise impact of the medium on quarkonia, it becomes necessary first to find the effective temperature of the quarkonia within the medium.

\subsubsection{Effective Temperature}
As mentioned, quarkonium states do not thermalize with the surrounding partonic medium due to their heavy mass and velocities. Therefore, the temperature of the quarkonia in its rest frame can be effectively described through the relativistic Doppler shift (RDS), which arises from the relative motion between the quarkonia and the thermal medium. The resulting effective temperature is quantified in terms of the relative velocity $v_{r}$ between the quarkonium states and the local fluid cell of the medium. The velocities of the medium and the quarkonium states are denoted by $v_{m}$ and $v_{Q(nl)}$, respectively. The latter is related to the quarkonium transverse momentum ($p_{\rm T}$) given as $v_{Q(nl)} = p_{\rm T}/E_{\rm T}$, where $E_{\rm T} = \sqrt{p_{\rm T}^{2} + M_{Q(nl)}^{2}}$ and $M_{Q(nl)}$ is the mass of the quarkonium states. The medium thermal velocity $v_{m}$  is obtained using the Maxwell-J\"{u}ttner distribution~\cite{Singh:2025xrd}. The RDS effect leads to an angle-dependent effective temperature experienced by the quarkonium, which is given as~\cite{Singh:2018wdt, Singh:2021evv, Singh:2025xrd};

 \begin{equation}
 T_{\text{eff}}(\tau, b, p_{\rm T}) = \frac{T(\tau,b) \sqrt{1 - |v_{r}|^{2}}}{1 - |v_{r}| \cos\theta},
 \label{tt}
 \end{equation}
where, $\theta$ denotes the angle between $v_{m}$ and $v_{Q(nl)}$. The medium temperature
$T(\tau,b)$ is obtained from the solution of Eqs.~(\ref{so1}) and (\ref{so2}). To obtain an
isotropic (angle-independent) representation of the effective temperature, Eq.~(\ref{tt}) is
averaged over the solid angle, yielding the mean effective temperature as;

 \begin{equation}
 T_{\text{eff}}(\tau, b, p_{\rm T}) = T(\tau,b)\frac{\sqrt{1 - |v_{r}|^{2}}}{2|v_{r}|}
 \ln\left(\frac{1 + |v_{r}|}{1 - |v_{r}|}\right)
 \end{equation}
The effective temperature of quarkonia encapsulates the modification of the local thermal
environment perceived by the moving quarkonium in a relativistically expanding medium.

\begin{figure}[ht!]
\includegraphics[scale = 0.4]{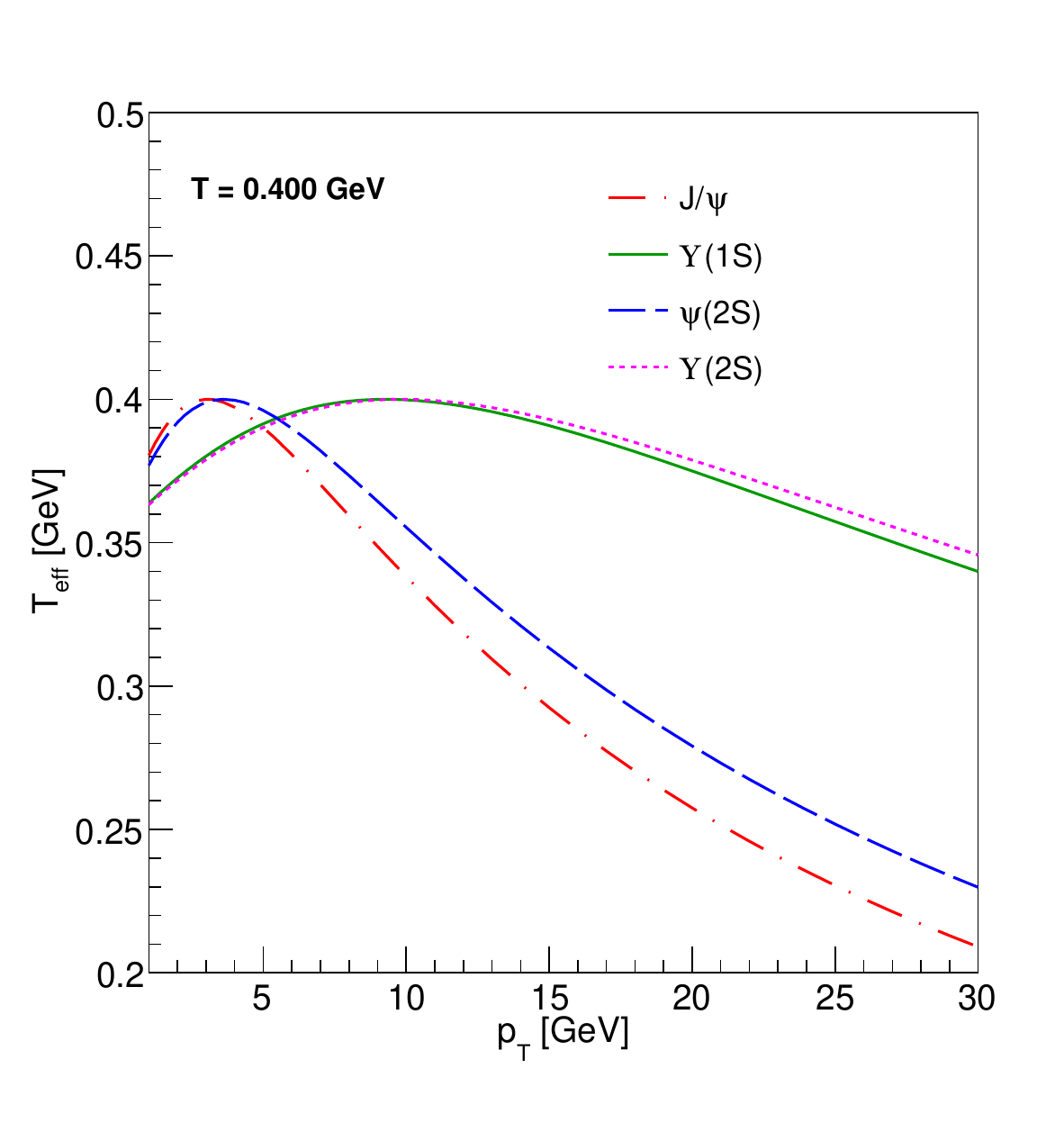}
\caption{(Color online) The effective temperature ($T_{\rm eff}$) as a function of $p_{\rm T}$ for $J/\psi$, $\psi$(2S), $\Upsilon$(1S), and $\Upsilon$(2S) states calculated at medium temperature $T$ = 0.400 GeV.}
\label{fig:Teff}
\end{figure}

The $T_{\rm eff}$ at $T = 400$ MeV as a function of $p_{\rm T}$ for different quarkonium states is
characterized in Fig.~\ref{fig:Teff}. The initial increase in $T_{\rm eff}$ at $p_{\rm T} \le
M_{Q(nl)}$ arises from the effective blue-shift induced when the quarkonium velocity is much smaller
than the medium flow velocity, $v_{Q} \ll v_m$. In this regime, the local thermal distribution
sampled by the quarkonium is shifted toward higher apparent temperatures due to the relative motion
between the probe and the expanding medium. At $v_{Q} \gg v_m$, corresponding to the $p_{\rm T}\gg
M_{Q(nl)}$, the kinematic red-shift dominates, reducing the apparent slope of the thermal spectrum
and consequently decreasing $T_{\rm eff}$ for both charmonium and bottomonium states. This
systematic behavior indicates that $T_{\rm eff}$ encodes the interplay between quarkonium kinematics
and medium flow, and therefore has a direct impact on the in-medium decay width, and eventually
influences the spin alignment of quarkonia as it depends on particle decay width (discussed in
Sec.~\ref{saq}).

\subsubsection{Quantum characteristic in presence of rotation}
The dynamics of a particle in a rotating medium can be described by a modified Hamiltonian that accounts for the coupling between the spin ($\bf{S}$) of the particle and the angular velocity $\boldsymbol{\omega}$ of the rotating medium. This interaction introduces rotational corrections to the kinetic and potential terms of the system, and the corresponding Hamiltonian can be expressed as~\cite{Anan, Gordan, Griffth},
\begin{align}
\mathcal{H} = \frac{1}{2m}\left(\boldsymbol{p}   - m\boldsymbol{\omega}\times\mathbf{r} \right)^2   - \frac{m}{2}(\boldsymbol{\omega}\times\mathbf{r})^2 -\boldsymbol{\omega}\cdot\mathbf{S}+ V(\boldsymbol{r})
\label{RotH}
\end{align}

For a two-body bound system such as quarkonium immersed in a rotating medium, Eq.~(\ref{RotH}) can be generalized as;

\begin{align}
&\mathcal{H} = \sum_{i=1,2}\left[\frac{1}{2m_i}(\boldsymbol{p}_i - m_i\boldsymbol{\omega}_i\times\mathbf{r}_i)^2 \right. \nonumber \\
& \left. - \frac{m_{i}}{2}(\boldsymbol{\omega_{i}}\times\mathbf{r}_{i})^2 - \boldsymbol{\omega_{i}}\cdot\mathbf{S_i}\right]  + V(\lvert \mathbf{r_1} - \mathbf{r_2} \rvert)
\label{RotH2}
\end{align}	

Here, the index $ i=1, 2$ refers to the heavy quark and its antiquark respectively. Applying the
standard reduction procedure for a two-body system to an effective one-body problem in the center of
the mass frame, the corresponding canonical momenta and position vectors can be defined as;

\begin{equation}
\left\{
\begin{array}{lcl}
\mathbf{P} = \mathbf{p_1} + \mathbf{p_2},  & &
\mathbf{p} = m_{\mu} \left(\dfrac{\mathbf{p_1}}{m_1} - \dfrac{\mathbf{p_2}}{m_2}\right) \\[3mm]
m_{\mu} = \dfrac{m_1\;m_2}{m_1+m_2},  &\quad & \\[3mm]
\mathbf{r_1} = \mathbf{R} +\dfrac{m_2}{m_1 + m_2}\mathbf{r}, & &
\mathbf{r_2} = \mathbf{R} -\dfrac{m_1}{m_1 + m_2}\mathbf{r}
\end{array}
\right.
\label{reduced}
\end{equation}

where $m_{\mu} = m_{q}/2$ represents the reduced mass of the heavy quark pair, with the charm and
bottom quark masses taken as $m_{c}=1.27$ GeV and $m_{b}=4.18$ GeV, respectively. Considering that
both the quark and antiquark experience the same rotation, i.e., $\boldsymbol{\omega}_1 =
\boldsymbol{\omega}_2 = \boldsymbol{\omega}$, the Hamiltonian simplifies as;

\begin{align}    
\mathcal{H} =&\; \frac{P^2}{2M} + \frac{p^2}{2m_{\mu}} - \mathbf{P}\cdot(\boldsymbol{\omega}\times\mathbf{R}) - \mathbf{p}\cdot(\boldsymbol{\omega}\times\mathbf{r})\nonumber \\
& - \boldsymbol{\omega}\cdot(\mathbf{S}_{1}+\mathbf{S}_{2})+ V(\lvert \mathbf{r} \rvert) 
\label{H2}	
\end{align}

Now, focusing on the internal motion in the reduced coordinate system, we obtain;

\begin{equation}
\mathcal{H} = \frac{p^2}{2m_{\mu}} - \mathbf{p}\cdot(\boldsymbol{\omega}\times\mathbf{r}) - \boldsymbol{\omega}\cdot(\mathbf{S}_{1}+\mathbf{S}_{2}) + V(\lvert \mathbf{r} \rvert) 
\label{H3}
\end{equation}
Taking rotation along the $z$-axis, i.e., $\boldsymbol{\omega} = \omega \hat{z}$, Eq.~(\ref{H3}) can be rewritten as;

\begin{equation}
\mathcal{H} = \frac{p^2}{2m_{\mu}} - \mathbf{p}\cdot(\boldsymbol{\omega}\times\mathbf{r}) - \omega (S_{1z}+S_{2z}) + V(\lvert \mathbf{r} \rvert) 
\label{H4}
\end{equation}

Using the operators $\mathbf{p} = - i \boldsymbol{\nabla}$, $\mathcal{H} = i\frac{\partial}{\partial t}$, $L_z = -i\frac{\partial}{\partial \phi}$, and $S_{1z}+S_{2z}=S_z$,  we can rewrite the Eq.~(\ref{H4}) as;

\begin{equation}
i\frac{\partial}{\partial t}= -\frac{1}{2m_{\mu}}\boldsymbol{\nabla}^2 - \omega (L_z + S_z) + V(\lvert \mathbf{r} \rvert) 
\label{NonIn}	
\end{equation}

By expressing the Laplacian operator in spherical coordinates and applying the method of separation of variables, the radial component of the Schr\"{o}dinger equation can be written as

\begin{align}
\frac{1}{r^2}\frac{d}{dr} \Bigl(r^2\frac{dR(r)}{dr}\Bigr)
+ 2m_{\mu} \Bigg[
(E-V(r)) - \frac{l(l+1)}{2 m_{\mu} r^2} \nonumber \\
- \omega(L_{z} + S_{z})
\Bigg] R(r) = 0
\label{shrr1}
\end{align}

Now, to include the effect of rotation explicitly in Schr\"{o}dinger equation, the angular velocity $\omega$ can be expressed in terms of the conserved circulation parameter $C$, defined as,

\begin{equation}
C = \oint \boldsymbol{v}\cdot \boldsymbol{dl}
\label{C}
\end{equation}

Using Stokes' theorem, which relates the line integral to the surface integral of vorticity, Eq.~(\ref{C}) can be rewritten as;

\begin{equation}
C =  \oint \boldsymbol{\nabla} \times \boldsymbol{v}  \cdot \boldsymbol{dS} = 2 \omega \pi r^2 
\label{C2}
\end{equation}

here, $\boldsymbol{\omega} = \frac{1}{2}\boldsymbol{\nabla} \times \boldsymbol{v}$ represents the non-relativistic vorticity. The rotational motion is treated classically and can be related to hydrodynamic vorticity quantities such as kinematic, thermal, temperature, and enthalpy vorticity.\\ 

Substituting the expression for $\omega$ in terms of $C$ into Eq.~(\ref{shrr1}) yields the modified radial Schr\"odinger equation in the rotating frame:

\begin{align}
\frac{1}{r^2}\frac{d}{dr}\left(r^2\frac{dR(r)}{dr}\right) + 2m_{\mu} \left[(E-V(r))-\frac{l(l+1)}{2 m_{\mu} r^2} \right. \nonumber \\
\left. - \frac{m_{j}C}{2\pi r^{2}} \right]R(r)=0	
\label{shrr2}
\end{align}
Here, $m_j$ is the magnetic quantum number that arises due to the coupling between $L_z$ and $S_z$, i.e., $L_z + S_z =  J_z$. For a vector meson like  quarkonium states $m_j = +1, 0, -1$.\\

We have numerically solved the Schr\"{o}dinger equation by incorporating the real part of the medium modified potential, $V(r, m_{D})$, as given by Eq.~(\ref{pot}) and obtained the eigen energies and wave functions for the corresponding quarkonium states. Furthermore, using these variables, we quantified the quarkonia dissociation mechanisms in the medium, which are discussed briefly in the following section.

\subsubsection{Collisional Damping}

The collisional energy loss of quarkonium in the QGP is commonly referred to as collisional damping, which arises due to the interactions of the bound heavy quark pair with the thermal partons in the medium. This effect leads to a finite in-medium width of the quarkonium state, reflecting its partial decoherence and eventual dissociation. The dissociation rate associated with this mechanism can be evaluated using the imaginary part of the potential, $V(r,m_{D})$, which encodes the medium-induced damping effect. The in-medium singlet complex potential for the heavy quark-antiquark bound state is given as~\cite{nendzig,Singh:2015eta}:

\begin{multline}
V(r,m_D) = \frac{\sigma}{m_D}(1 - e^{-m_D\,r}) - \alpha_{\rm eff} \left ( m_D
+ \frac{e^{-m_D\,r}}{r} \right )\\
- i\alpha_{\rm eff} {T_{\rm eff}} \int_0^\infty
\frac{2\,z\,dz}{(1+z^2)^2} \left ( 1 - \frac{\sin(m_D\,r\,z)}{m_D\,r\,z} \right)
\label{pot}
\end{multline}

In Eq.~(\ref{pot}), the first two terms on the right-hand side correspond to the screened string and Coulombic interactions, respectively. The third term represents the imaginary part of the potential, which accounts for the collisional or Landau damping of the low-frequency gauge fields that mediate interactions between quark-antiquark pairs. In Eq.~(\ref{pot}),  $\sigma$ represents the string tension between the quark-antiquark pair and taken as $\sigma =$ 0.192 GeV$^{2}$. Further, $m_D$ is the Debye screening mass, given as;

\begin{equation}
m_{D} = T_{\rm eff} \sqrt{4\pi\alpha_{s}^{T} \left( \frac{N_c}{3} +\frac{N_f}{6} \right)}
\label{mDeB}
\end{equation}
here,  $N_c = 3$,  $N_f = 3$  and $\alpha_s^T$ is the strong coupling constant at the hard scale obtained under the Hard Thermal Loop (HTL) limit. In general, the leading order strong running coupling constant $\alpha_{s}$ is defined as;
\begin{equation}
\alpha_{s}(\mu)  = \frac{4\pi}{\beta_{0}\ln\left(\frac{\mu^{2}}{\Lambda_{\overline{MS}}^2}\right)} 
\label{alpha}
\end{equation}

where $\beta_{0}= 11 - \frac{2}{3}N_f$, $\Lambda_{\overline{MS}}$ = 0.176 GeV, and $\mu$ is
renormalization scale. The $\alpha_s^T$ is obtained by scaling at the HTL limit ($\mu$  = $2 \pi
T$), i.e. $\alpha_{s}^{T} = \alpha_{s}(2\pi T)$. As it can be seen from Eq.~(\ref{alpha}), the
$\alpha_{s}^{T}$ depends on medium temperature. At a particular temperature, say T = 400 MeV,
$\alpha_{s}^{T} \simeq$ 0.26. Likewise, the coupling constant at the soft scale, $\alpha_s^s$ and
ultra soft scale $\alpha_s^u$  are determined by evaluating the running coupling at the
characteristic scales associated with heavy-quark dynamics; $\alpha_{s}^{s} = \alpha_{s}
(m_{q}\alpha_{s}/2)$ and $\alpha_{s}^{u} = \alpha_{s}(m_{q}\alpha_{s}^{2}/2)$. Here, $m_q$ is the
mass of the charm/bottom quark. Following Eq.~(\ref{pot}), effective coupling entering the
potential is defined in terms of soft-scale coupling as
$\alpha_{\mathrm{eff}}=\frac{4}{3}\alpha_s^{s}$. For the chosen set of parameters, we obtain
$\alpha_s^{s}\simeq  0.33$ for bottom quark, and $\alpha^{s}_{s} \simeq 0.5$ for charm quark, which
fixes the numerical value of $\alpha_{\mathrm{eff}}$, respectively.\\

The collisional damping width, $\Gamma_{\rm damp,nl}$, quantifies the in-medium decay rate of quarkonium due to the imaginary part of the potential. It is computed using first-order perturbation theory as an expectation value of $\mathrm{Im}(V)$ with the radial wave function of the bound state:\\

\begin{equation}
\Gamma_{{\rm damp} ,nlm}(\tau,p_{\rm T}) = \int[g_{nlm}(r)^{\dagger} \left [ \rm Im(V)\right] g_{nlm}(r)] dr,
\label{cold}
\end{equation}
here $g_{nlm}(r)$ is the quarkonium singlet wave function. The spin-dependent wave functions for $J/\psi$, $\psi$(2S), $\Upsilon$(1S) and $\Upsilon$(2S)  are obtained by solving the Schr\"{o}dinger equation 
(Eq.~(\ref{shrr2})) in the presence of rotation, as described in the preceding section.

\subsubsection{Gluonic Dissociation}
The gluon-induced dissociation (or gluonic dissociation) of quarkonia in the QGP medium is another
dissociation mechanism. This  process arises when a thermal gluon interacts inelastically with a
color-singlet quarkonium state, leading to a transition to a color-octet state, which subsequently
breaks apart. The corresponding dissociation cross-section depends on the gluon energy spectrum and
temperature profile of the medium. Following the formalism introduced by Peskin and
Bhanot~\cite{Peskin:1979va, Bhanot:1979vb} and extended to finite-temperature
QCD~\cite{Grandchamp:2001pf, Brambilla:2008cx, Singh:2018wdt, Singh:2021evv}, the gluon-dissociation
cross section for a quarkonium state is given by,

\begin{multline}
\sigma_{{\rm d},nlm}(E_g) = \frac{\pi^2\alpha_s^u E_g}{N_c^2}\sqrt{\frac{{m_{q}}}{E_g
+ E_{nlm}}}\\
\;\;\;\;\;\times \left(\frac{l|J_{nlm}^{q,l-1}|^2 +
(l+1)|J_{nlm}^{q,l+1}|^2}{2l+1} \right)
\end{multline}
where,  $\alpha_{s}^{u} \simeq  0.76$ for bottom quark, and, $\alpha_{s}^{u} \simeq 0.87$ for charm quark. Here, $E_g$ is the incident gluon energy, and $E_{nlm}$ denotes the energy eigenvalue of the 
quarkonium state. The $g_{nlm}(r)$ 
and $h_{ql^{\prime}}(r)$ are the color-singlet and color-octet wave functions with orbital angular momenta $l$ and $l^{\prime}$, respectively. The octet 
wave function $h_{\rm ql^{\prime}}(r)$ has been obtained by solving the Schr\"{o}dinger equation with the octet potential 
$V_{8} = \alpha_{\rm eff}/8r$. For a given value of $n$ and $l$, the color-octet wave function and the corresponding energy eigenvalue are identical for all three spin projection states $m$. The value of $q$ is determined by using the conservation of energy, $q = \sqrt{m_{q}(E_{g}+E_{nlm})}$.\\

\noindent The probability density $J_{nlm}^{ql^{\prime}}$ encodes the overlap between the initial bound
and final continuum states and is computed using the in medium  singlet and octet wave functions,
\begin{equation}
 J_{nlm}^{ql^{\prime}} = \int_0^\infty dr\; r\; g^*_{nlm}(r)\;h_{ql^{\prime}}(r)
\end{equation}

As quarkonium moves with a finite velocity relative to the medium, the gluon distribution in the
quarkonium rest frame gets modified due to the relativistic Doppler shift of gluon energies. Under
a Lorentz transformation, the gluon energy in the quarkonium rest frame is modified as $E_g' =
\gamma E_g (1 + v_{Q}\cos\theta)$, where $\gamma$ is the Lorentz factor, and $\theta$ is the angle
between the quarkonium velocity vector and the incoming gluon momentum. Consequently, the modified
gluon distribution function in the quarkonium frame is expressed as,

\begin{equation}
f_g = \left[\exp\!\left(\frac{E_g}{T_{\text{eff}}}\,\gamma(1 + v_{Q}\cos\theta)\right) - 1\right]^{-1}.
\end{equation}

\noindent This modified distribution plays an essential role in determining the gluonic dissociation rate and, consequently, the momentum-dependent dissociation of quarkonia in the QGP. The thermal averaged gluonic dissociation, $\Gamma_{{\rm gd}, nlm}$, can then be obtained by folding the cross section with this modified Bose–Einstein distribution of thermal gluons in the particle rest frame~\cite{Singh:2018wdt, Singh:2021evv}, as given by;

\begin{equation}
\Gamma_{{\rm gd}, nlm}(\tau,p_{\rm T},b) = \frac{g_d}{4\pi^2} \int_{0}^{\infty}
\int_{0}^{\pi} \frac{dp_g\,d\theta\,\sin\theta\,p_g^2
\sigma_{{\rm d},nlm}(E_g)}{e^{ \{\frac{\gamma E_{g}}{T_{\rm eff}}(1 +
v_{Q}\cos\theta)\}} - 1}
\label{glud}
\end{equation}
here, $p_T$ is the transverse momentum of the quarkonium, and $g_d = 16$ represents the degeneracy factor for the gluons.\\ 

Furthermore, we have obtained the net decay width $\Gamma_{\rm D}$ for quarkonia in-medium via
taking the sum over Eq.~(\ref{cold}), and  (\ref{glud}), given as;

\begin{equation}
      \Gamma_{{\rm D}, nlm} = \Gamma_{{\rm damp},nlm} + \Gamma_{{\rm gd},nlm}
      \label{gammaD}
\end{equation}

\noindent
The combined effects of collisional damping and gluonic dissociation are subsequently incorporated into the computation of the spin alignment observable $\rho_{00}$, allowing us to explore the role of medium-induced spin–vorticity coupling of quarkonia within a rotating QGP. 

\subsection{Spin alignment of quarkonia}
\label{saq}

The dissociation probability of $m^{\rm th}$ spin state in the presence of rotation is given by;

\begin{equation}
    {\mathcal{P}}_{m} = \exp\left[ -\bigintssss_{\tau_0}^{\tau_f} \Gamma_{{\rm D}, nlm} (\tau, p_{\rm T})\; d\tau  \right].
\end{equation}

Here $m$ takes on the values $0, +1, -1$, representing the three distinct spin projections
corresponding to the states 0, $+$1, and $-1$, respectively. The $\tau_0$ and $\tau_f$ are the
proper times corresponding to the initial temperature $T_{0}$ and freeze-out temperature $T_{f}$
for the QGP medium. The dissociation rate, denoted as $\Gamma_{{\rm D}, nlm}$ is intricately linked
to both the rotational dynamics and the specific spin projection states, $m$. \\

The spin alignment of vector mesons is measured via the elements of the spin density matrix $\rho_{m, m'}$, where $m$ and $m'$ denote the spin projections along a chosen quantization axis. For a vector meson with spin 1, the spin density matrix is defined as;
\begin{equation}
\rho_{m,m^{\prime}} = \frac{\langle A_m A^{*}_{m^{\prime}}\rangle}{\sum \limits^{k = +1}_{k =-1} \langle |A_k|^{2} \rangle}
\label{rho00}    
\end{equation}
where $A_{m}$ denotes the production amplitude for the spin projection $m$, and the denominator ensures proper normalization. The density matrix satisfies the normalization condition
\begin{equation}
{\rm Tr(\rho)} = \sum \limits^{m = +1}_{m =-1}  \rho_{m,m} = 1
\label{rho00}    
\end{equation}

The diagonal elements of the spin density matrix are given by

\begin{equation}
\rho_{m,m} = \frac{\langle |A_m|^{2}\rangle}{\sum \limits^{k = +1}_{k =-1} \langle |A_k|^{2} \rangle}
\label{rho00}    
\end{equation}

The production amplitude $A_{m}$ for the spin projection $m$, is related to the dissociation probability ${\mathcal{P}}_{m}$ via the relation

\begin{equation}
{\mathcal{P}}_{m} \equiv \langle |A_m|^{2}\rangle
\label{rho00}    
\end{equation}
The $00^{\rm th}$ element of the $3 \times 3$ spin density matrix is defined in terms of the $\Gamma_{{\rm D}, nlm}$;
\begin{equation}
\rho_{00} = \frac{{\mathcal{P}}_0}{{\mathcal{P}}_{+} + {\mathcal{P}}_{0} + {\mathcal{P}}_{-}}
\label{rho00}    
\end{equation}

\section{Results and Discussions}
\label{res}

\begin{figure*}[!htbp]
\centering
\includegraphics[scale = 0.4]{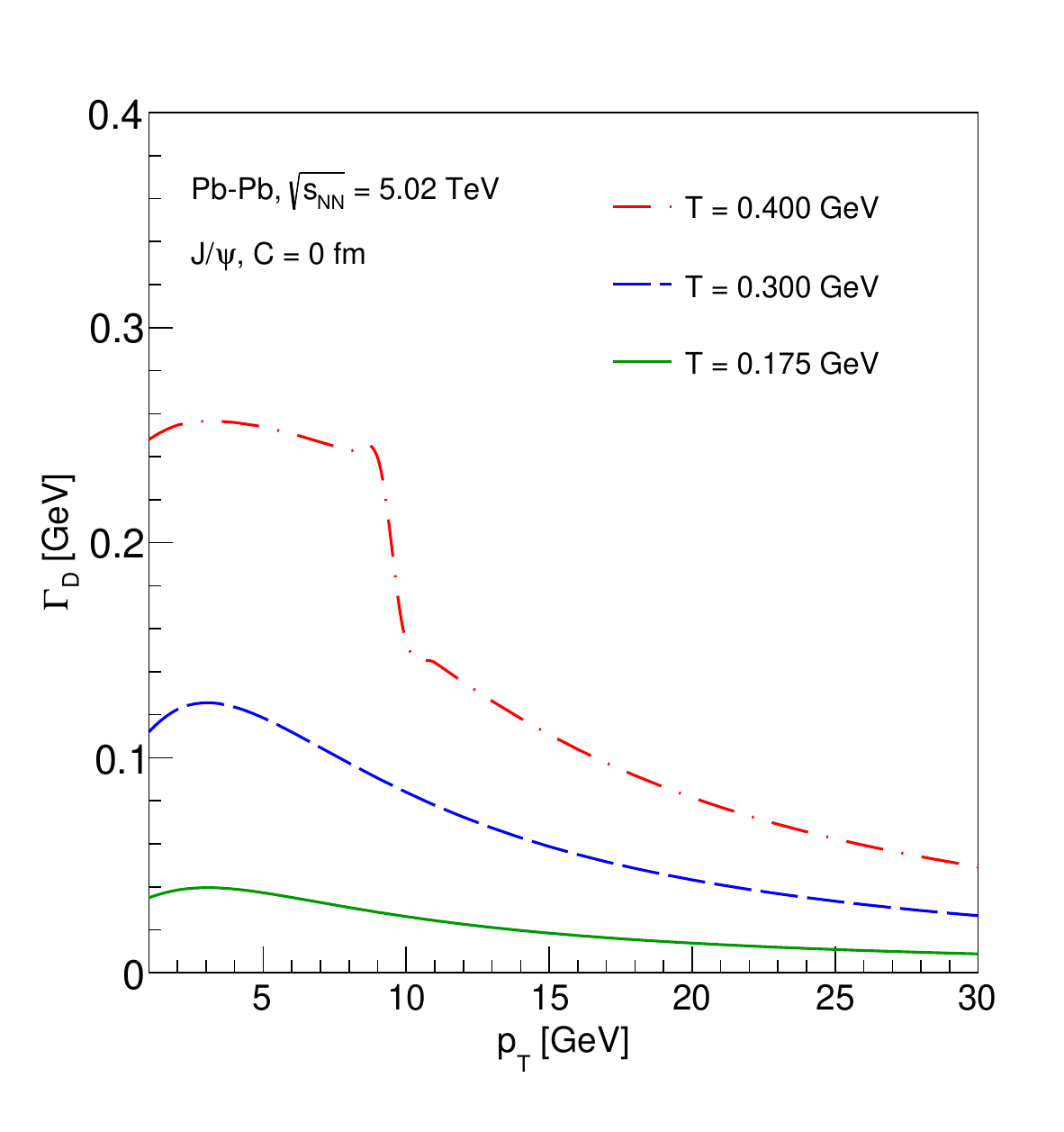}
\includegraphics[scale = 0.4]{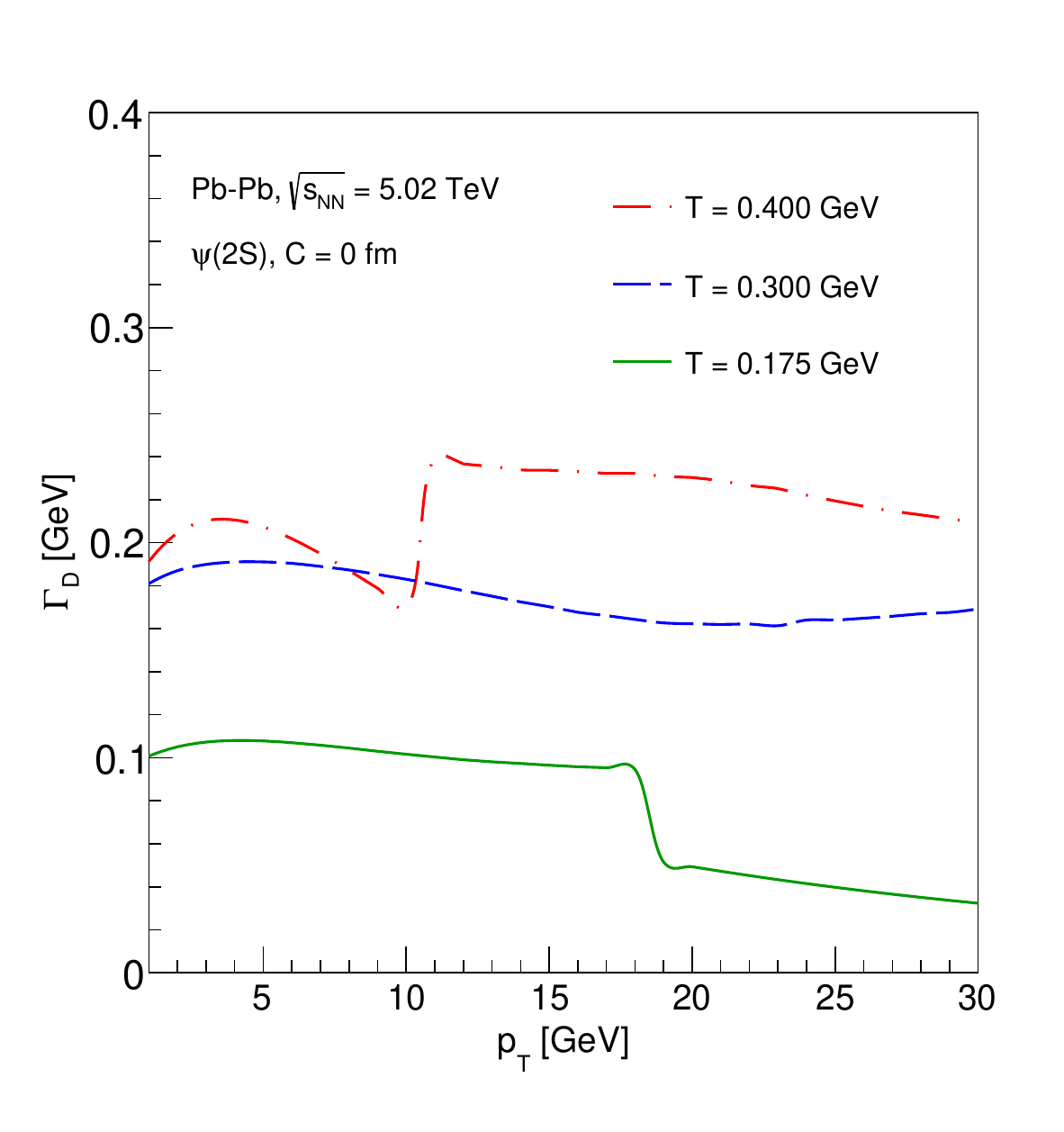}
\includegraphics[scale = 0.4]{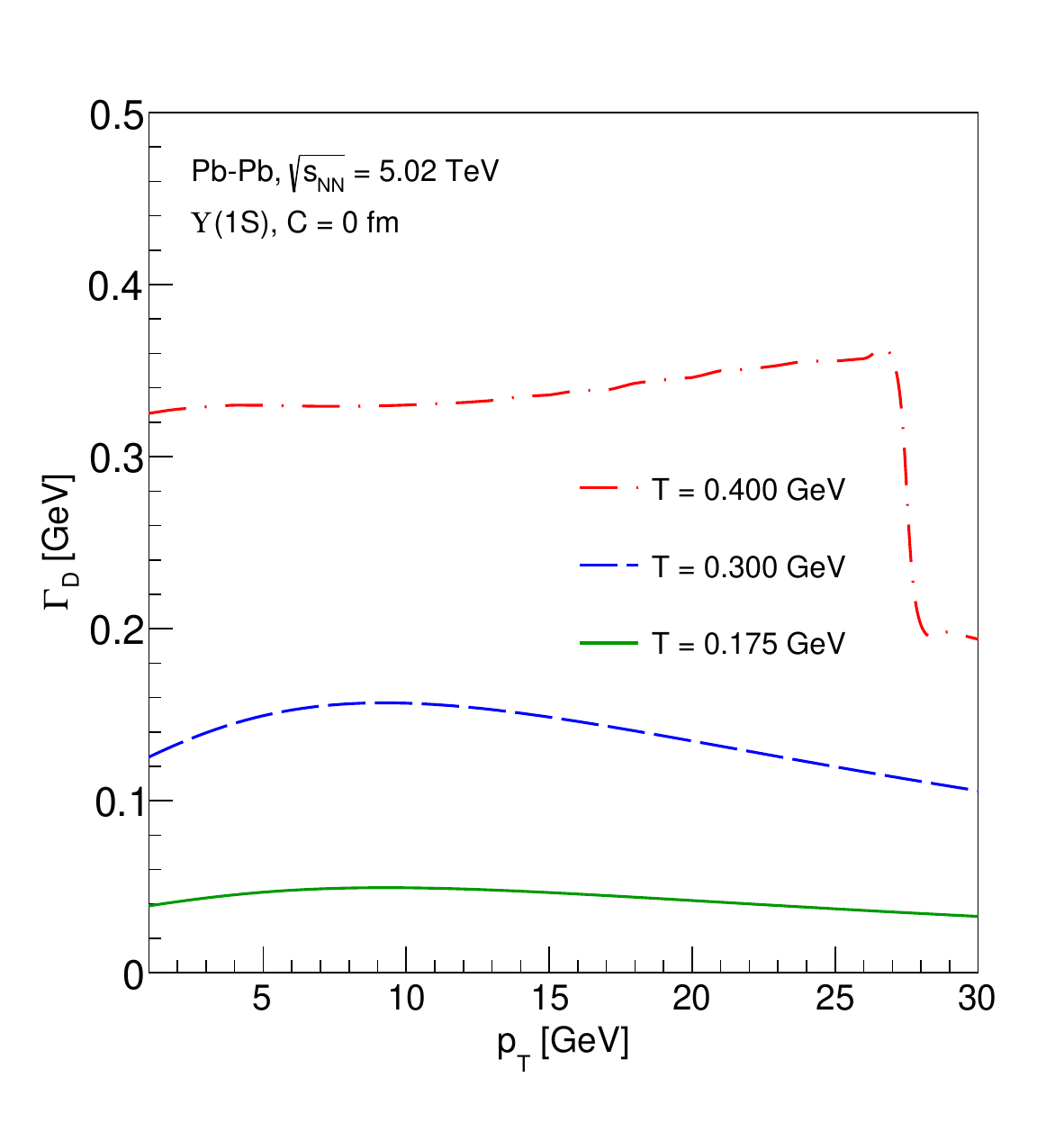}
\includegraphics[scale = 0.4]{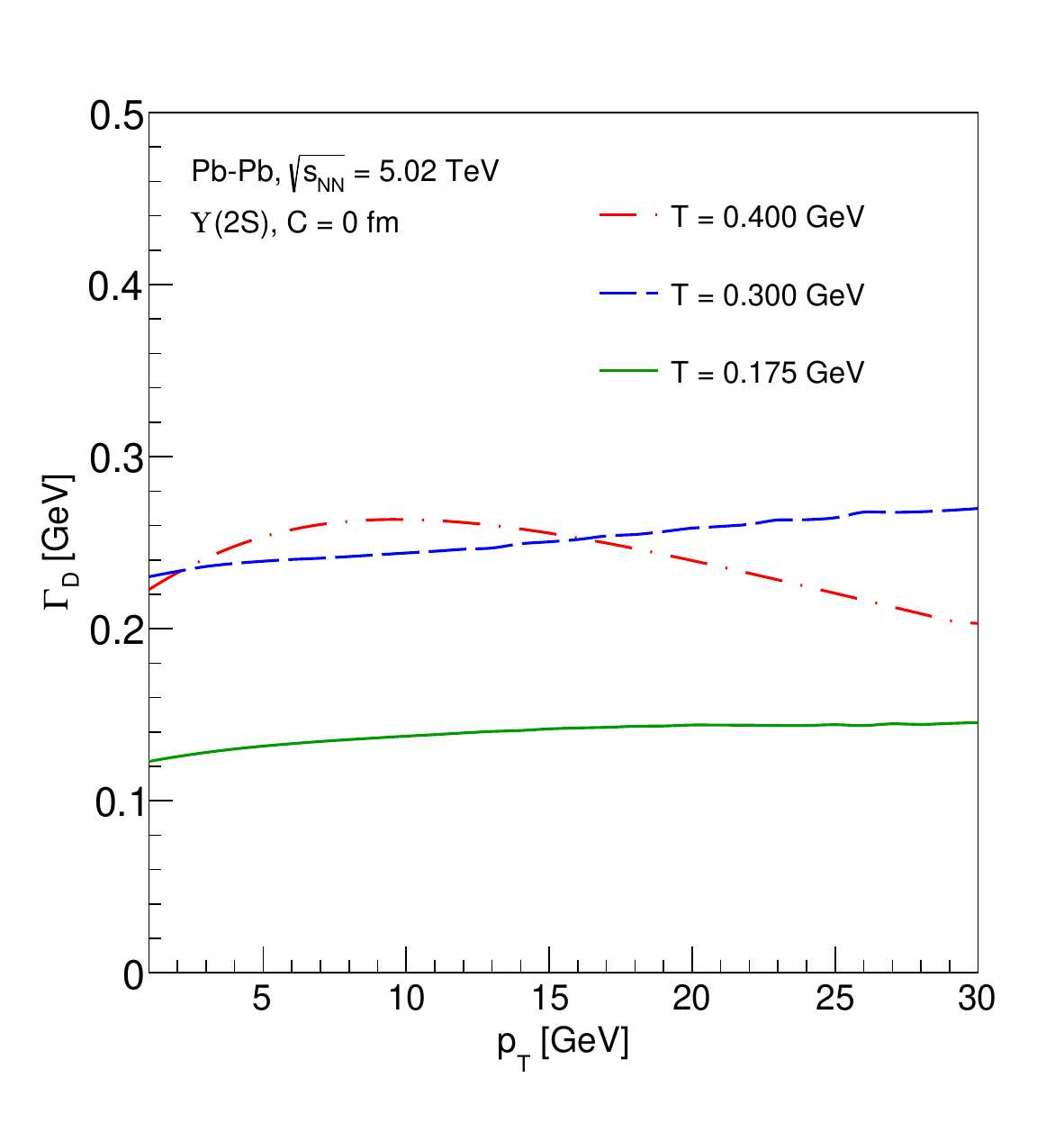}
\caption{The total decay width $\Gamma_{\rm D}$ as a function of transverse momentum ($p_{\rm T}$)
for $J/\psi$ (upper left), $\psi$(2S) (upper right),
$\Upsilon$(1S) (lower left), and $\Upsilon$(2S) (lower right) in Pb--Pb collisions at $\sqrt{s_{\rm
NN}}$ = 5.02 TeV. Results are presented for three different temperatures, such as $T$ = 0.400 GeV
(red dotted dashed line),  $T$ = 0.300 GeV (blue dashed line), and  $T$ = 0.175 GeV (green solid
line), with the circulation parameter $C = 0$ fm.}
\label{fig:GammaDwoC}
\end{figure*}

Quarkonium suppression and spin alignment provide complementary insights into the interaction
between heavy quark–antiquark pairs and the quark–gluon plasma. Following that, we evaluated how
different dissociation processes modify the spin alignment factor $\rho_{00}$ for the $J/\psi$,
$\psi$(2S), $\Upsilon$(1S), and $\Upsilon$(2S) states, as defined in Eq.~(\ref{rho00}). Here,
results are obtained by numerically solving the Schr\"{o}dinger equation, which provides
temperature, vorticity, and spin-projection-dependent eigenvalues and wave functions. The imaginary
component of the potential, together with the quarkonium wave function, determines the collisional
damping width, whereas the binding energy inferred from the eigenvalue governs the
gluon-dissociation width. As reported in the earlier findings~\cite{Singh:2015eta, Singh:2025xrd},
the collisional damping mechanism dominates over gluonic dissociation across the thermal evolution
of the medium.\\

\begin{figure*}[htbp]
\centering
\includegraphics[scale = 0.4]{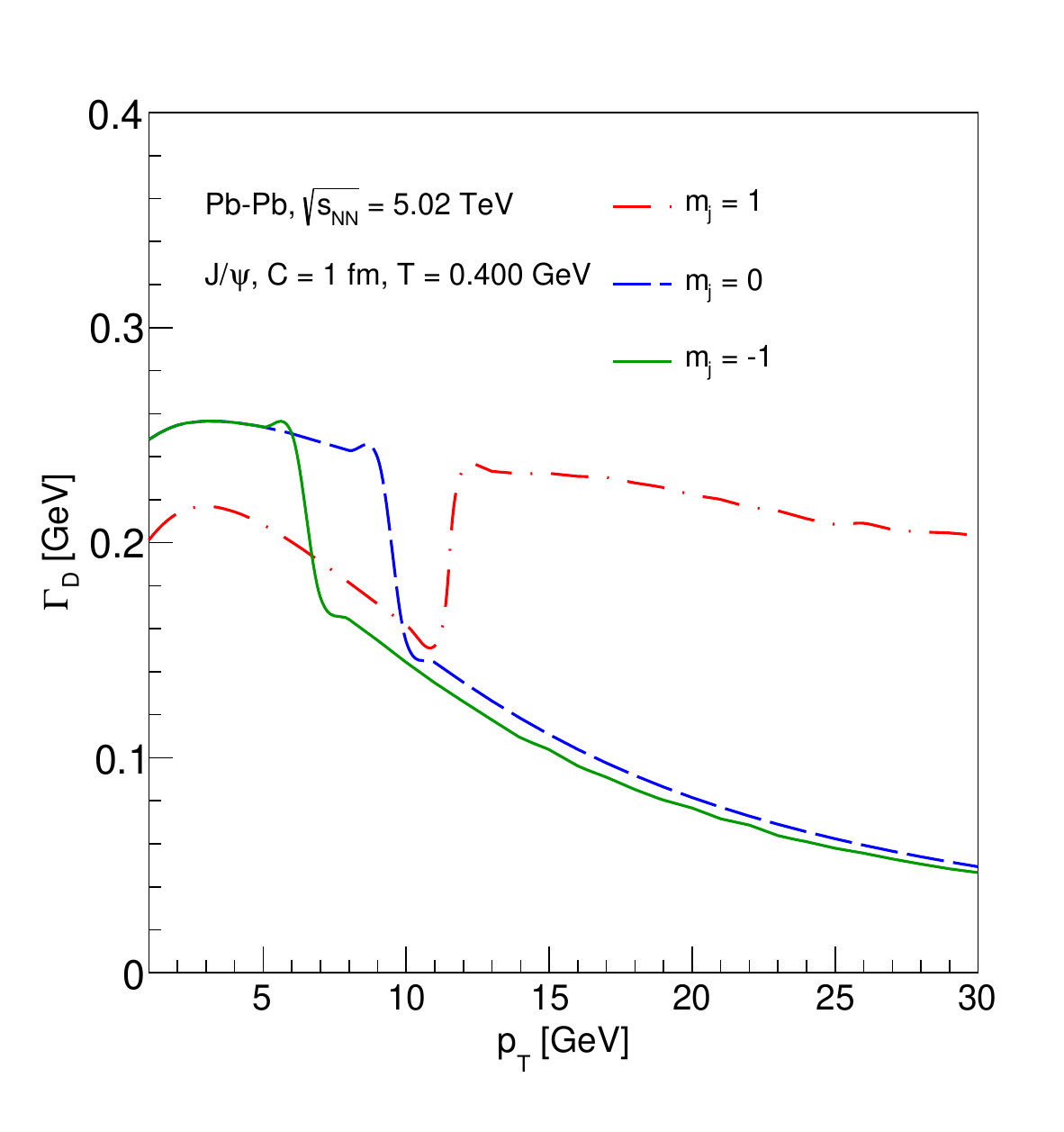}
\includegraphics[scale = 0.4]{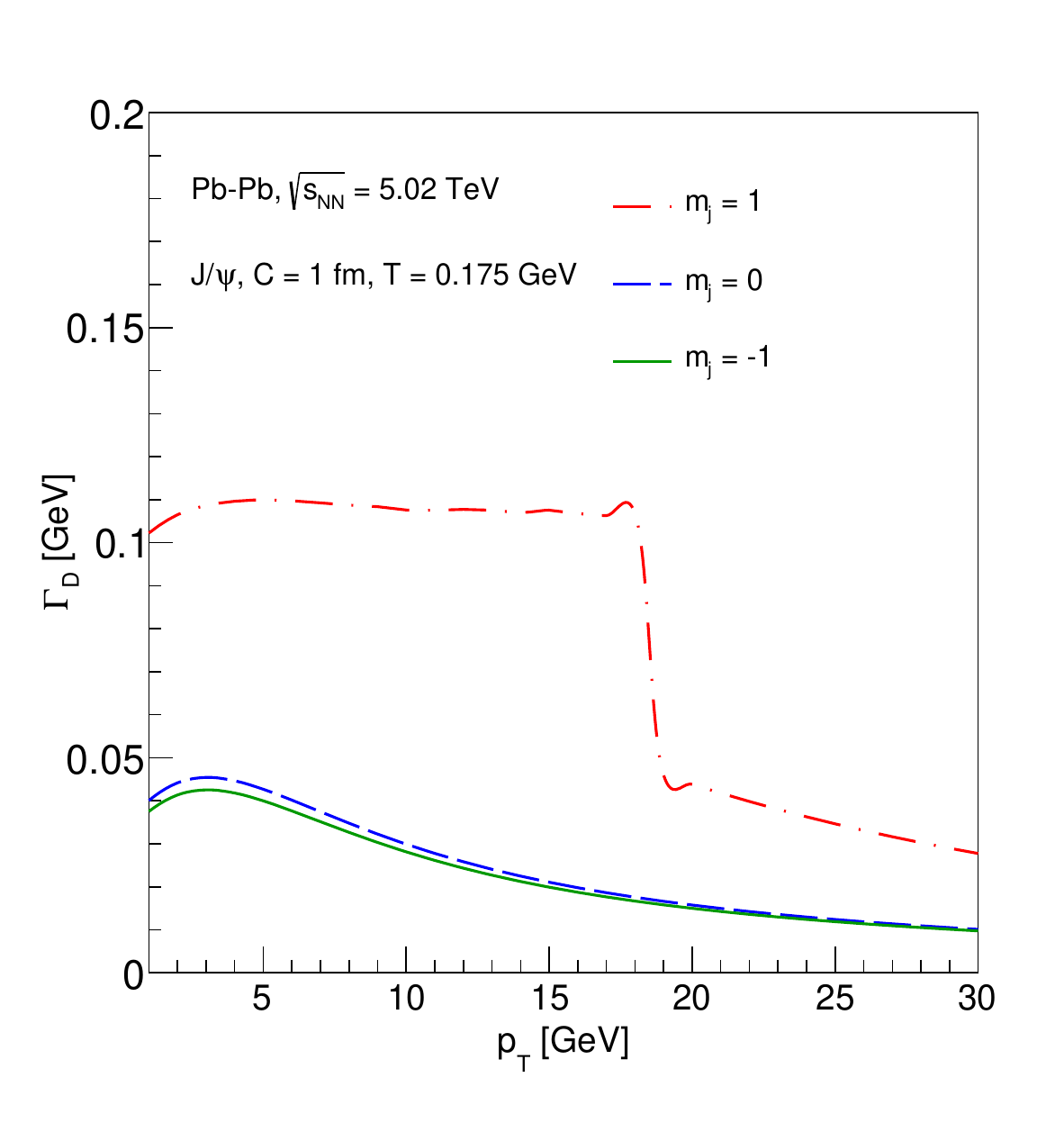}
\includegraphics[scale = 0.4]{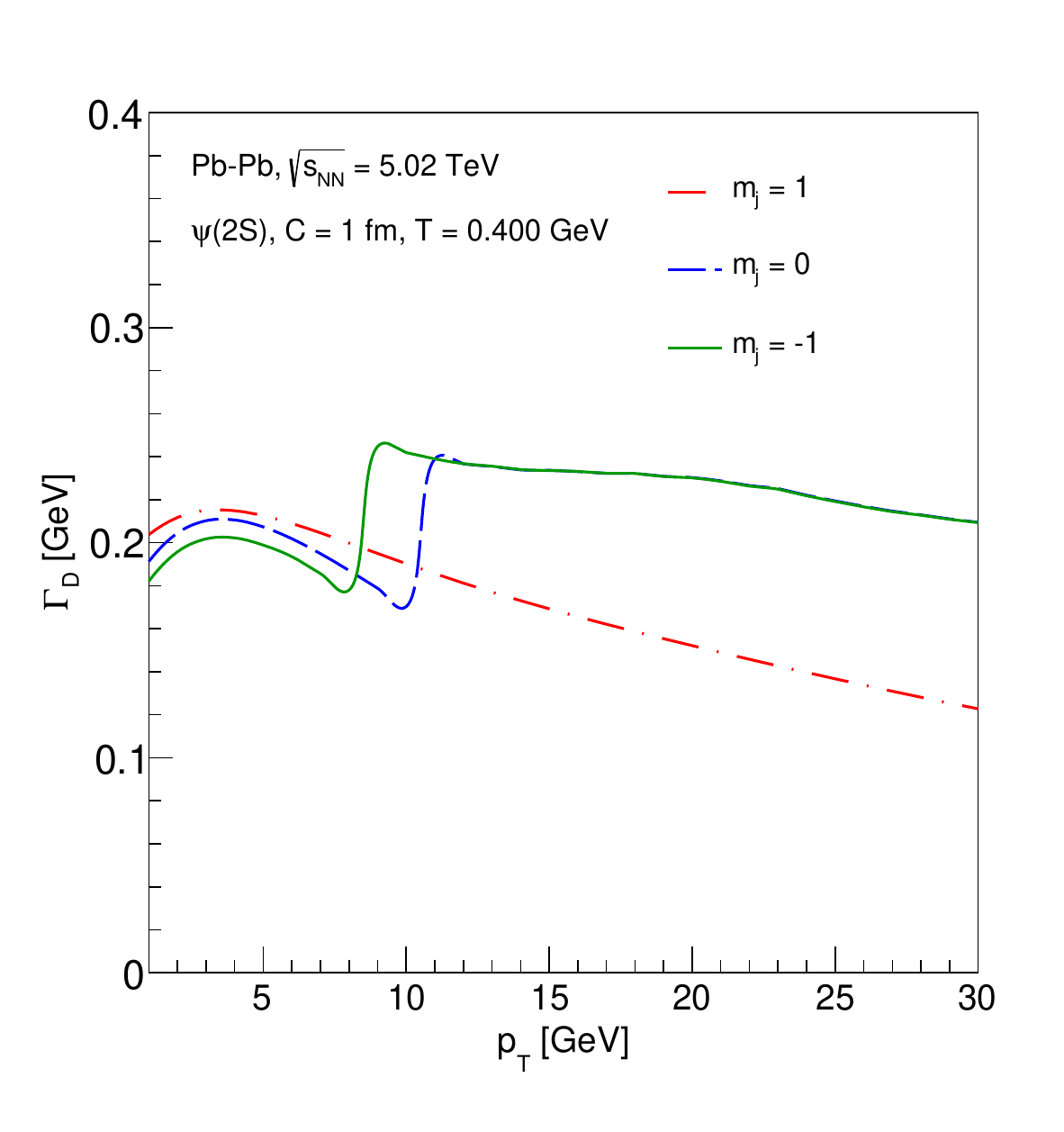}
\includegraphics[scale = 0.4]{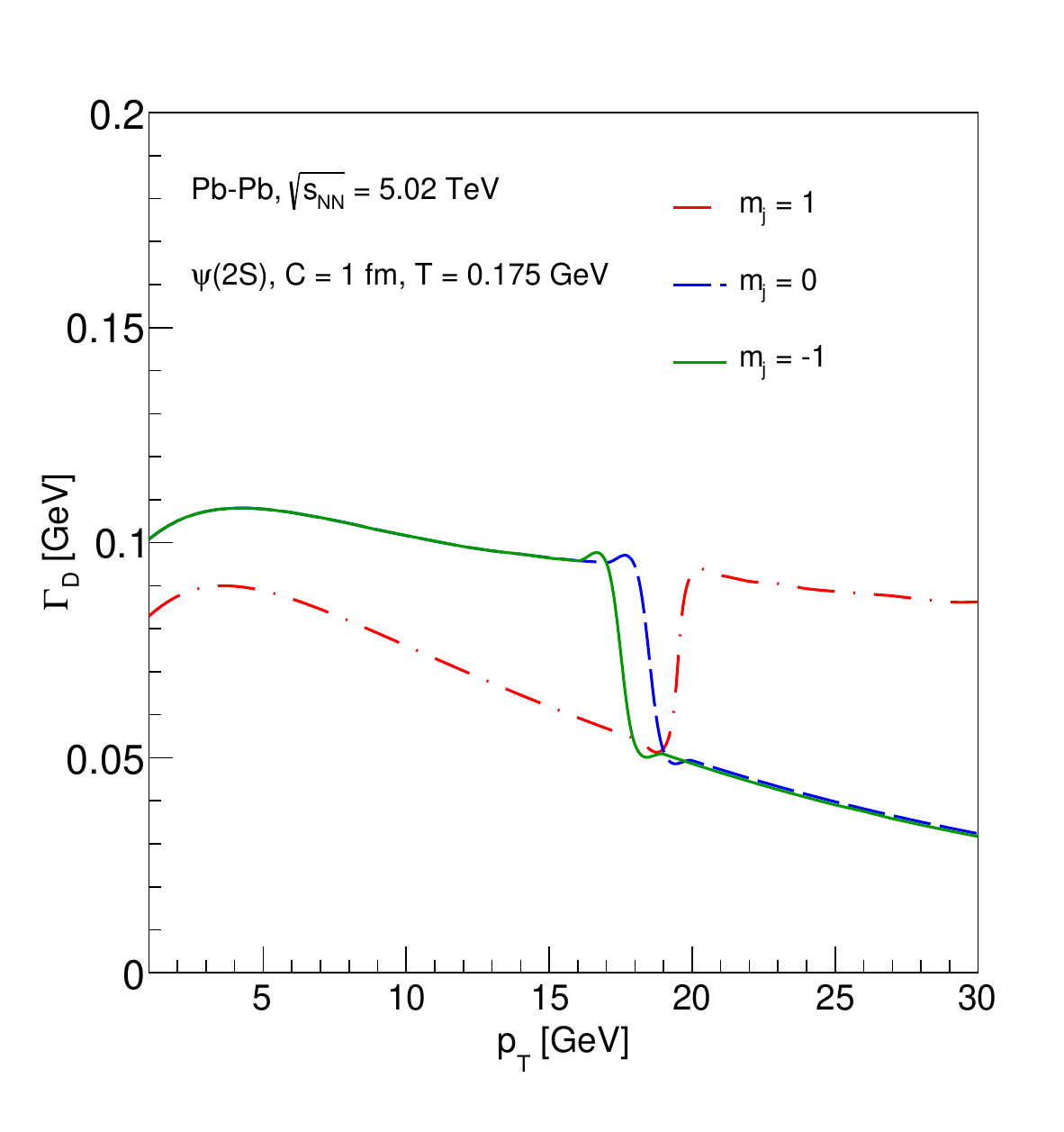}
\caption{The total decay width $\Gamma_{\rm D}$ as a function of transverse momentum ($p_{\rm T}$)
for $J/\psi$ (upper), and $\psi$(2S) (lower) in Pb--Pb collisions at $\sqrt{s_{\rm NN}}$ = 5.02 TeV
for three different spin projections $m_{j}=$ 1 (red dotted dashed line), $m_{j}=$ 1 (blue dashed
line), $m_{j}=$ 1 (green solid line) states with the circulation parameter $C = 1$ fm. The upper
(lower) left and right panels present the $J/\psi$ ($\psi$(2S)) decay width for two different
temperatures: $T$ = 0.400 GeV and 0.175 GeV, respectively.}
\label{fig:GammaDwCchar}
\end{figure*}

\begin{figure*}[htbp]
\centering
\includegraphics[scale = 0.4]{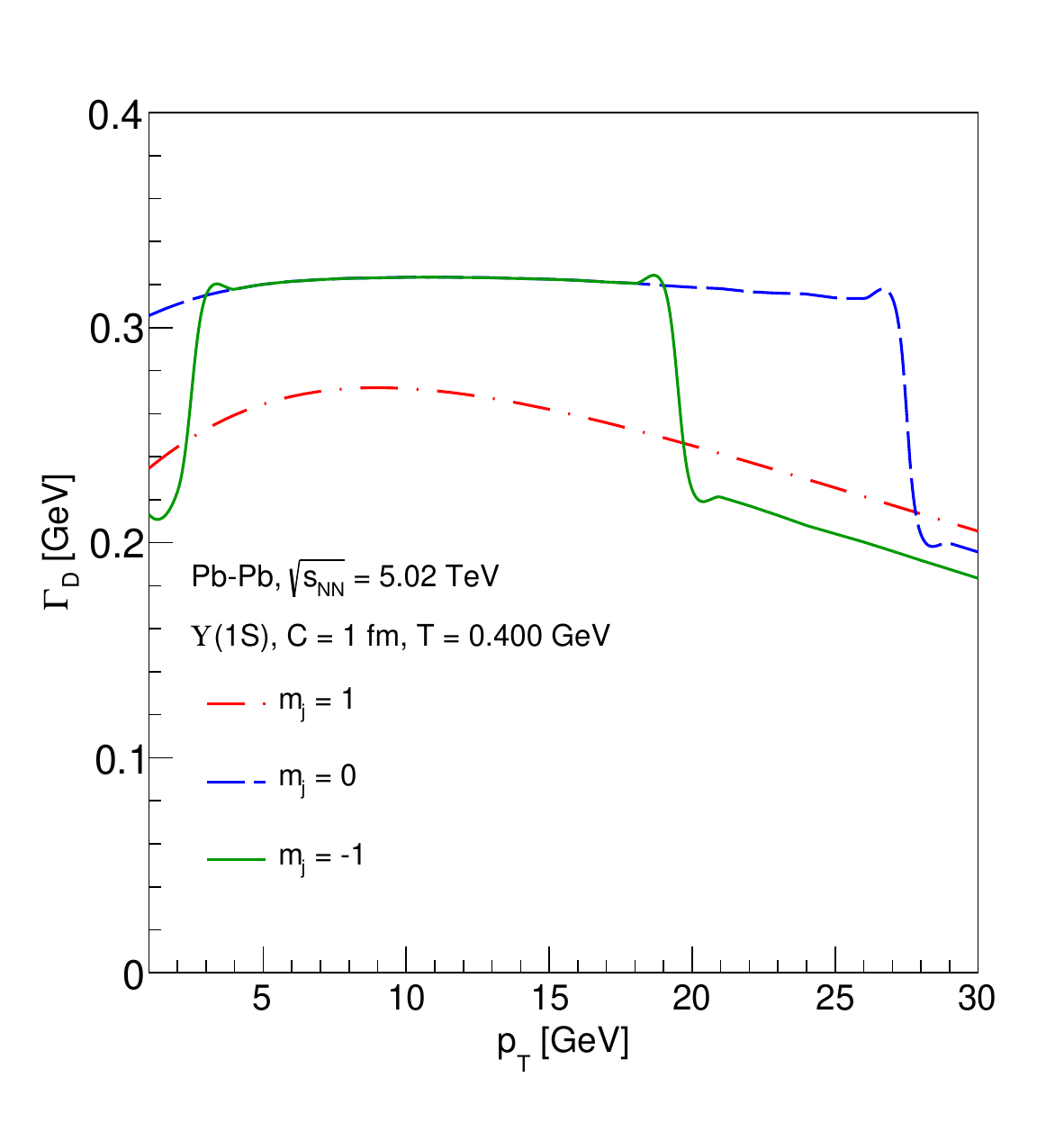}
\includegraphics[scale = 0.4]{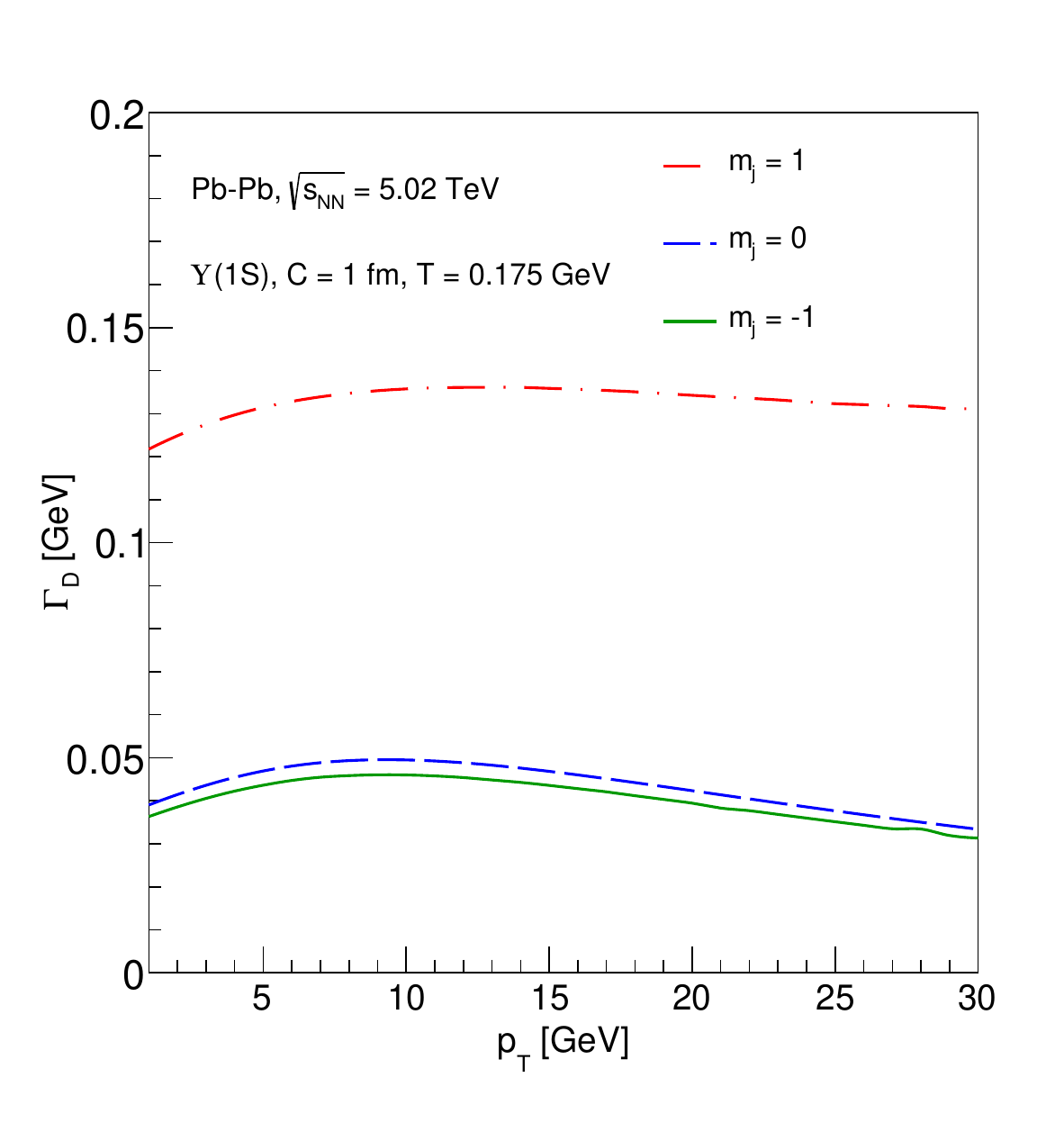}
\includegraphics[scale = 0.4]{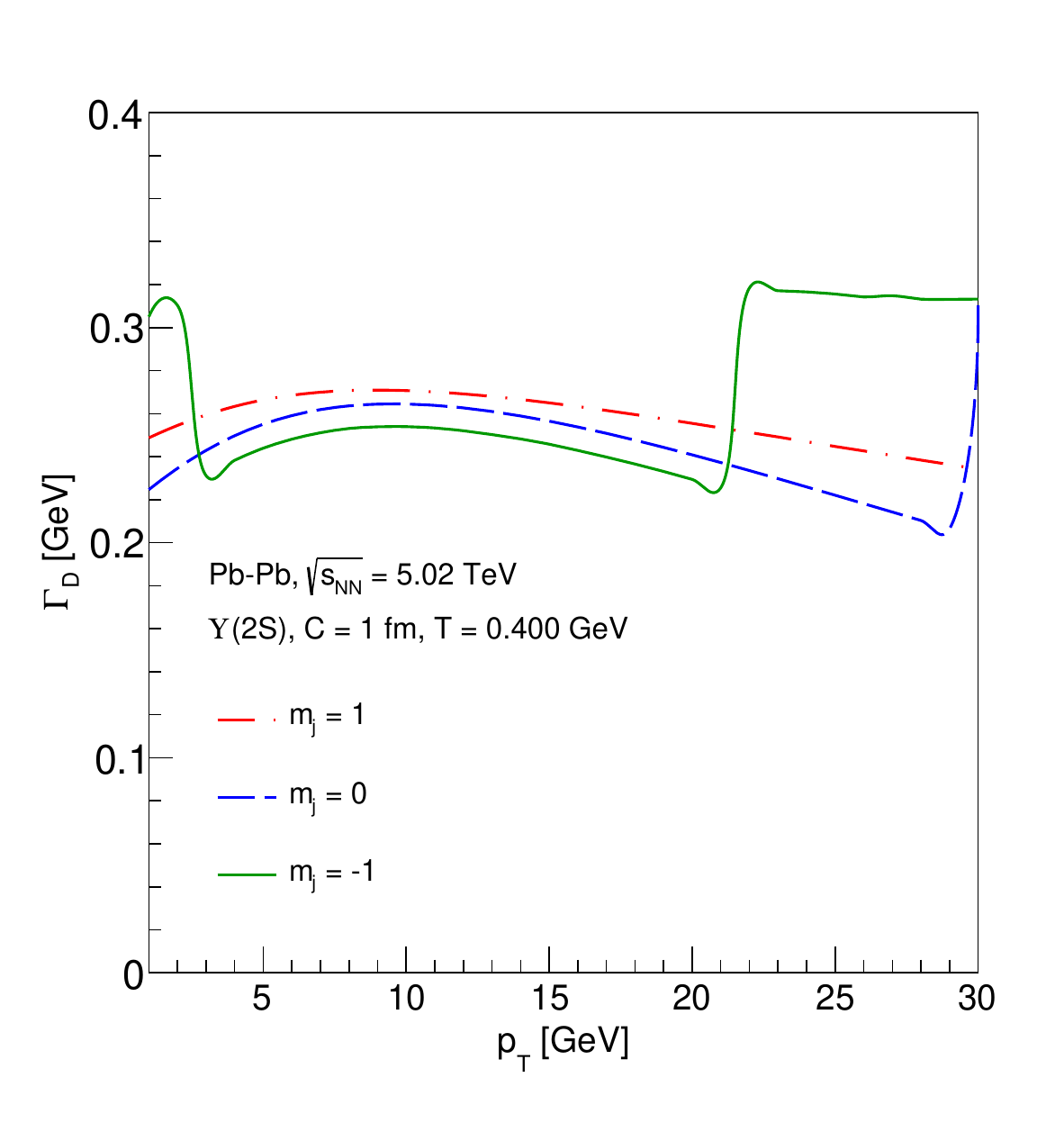}
\includegraphics[scale = 0.4]{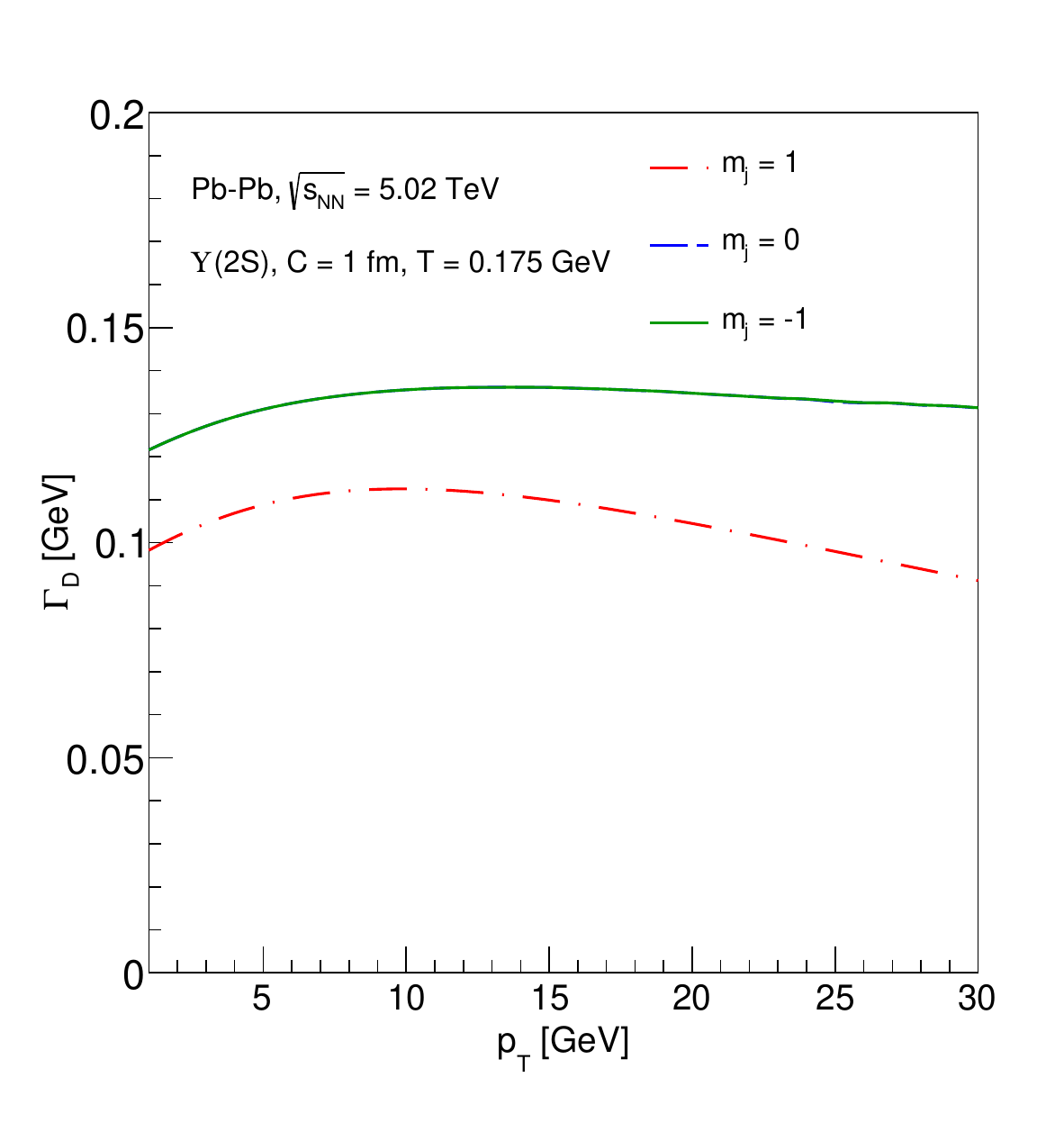}
\caption{The total decay width $\Gamma_{\rm D}$ as a function of transverse momentum ($p_{\rm T}$)
for $\Upsilon$(1S) (upper), and $\Upsilon$(2S) (lower) in Pb--Pb collisions at $\sqrt{s_{\rm NN}}$
= 5.02 TeV for three different spin projections $m_{j}=$ 1 (red dotted dashed line), $m_{j}=$ 1
(blue dashed line), $m_{j}=$ 1 (green solid line) states with the circulation parameter $C = 1$ fm.
The upper (lower) left and right panels present the $\Upsilon$(1S) ($\Upsilon$(2S)) decay width for
two different temperatures: $T$ = 0.400 GeV and 0.175 GeV, respectively.}
\label{fig:GammaDwCbott}
\end{figure*}

The decay width varies with both the magnitude and the direction of the vorticity field (quantified
in terms of the circulation parameter $C$) relative to the quarkonium spin projection. Along with
that, it also strongly depends on the transverse momentum ($p_{\rm T}$) and effective temperature
($T_{\rm eff}$) of the particle. It is noteworthy to mention that characteristics of $\Gamma_{\rm
D}$ as a function of $p_{\rm T}$ play a crucial role in shaping alignment in the final particle
distributions.\\

\begin{figure*}[htbp]
\centering
\includegraphics[scale = 0.4]{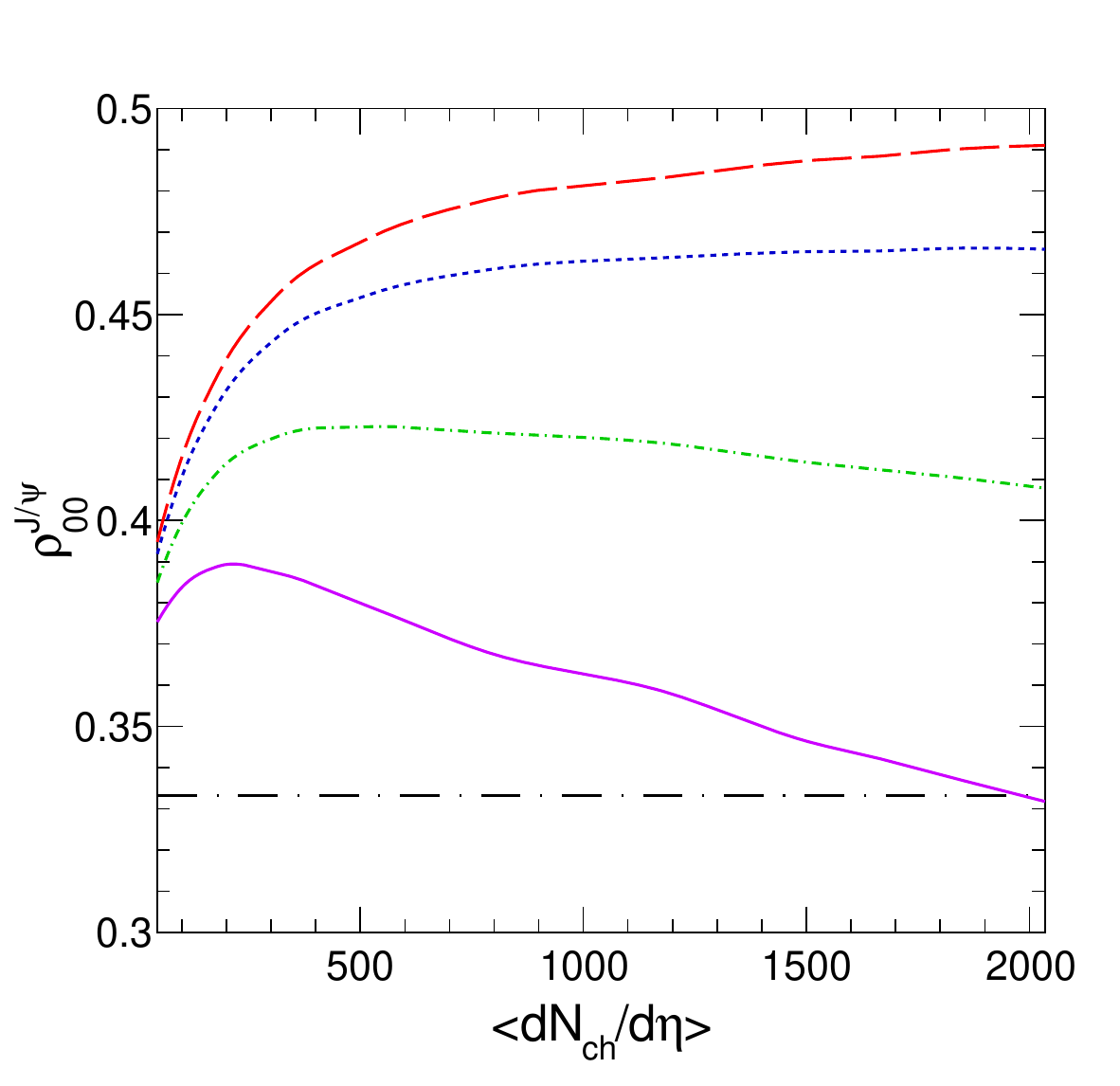}
\includegraphics[scale = 0.4]{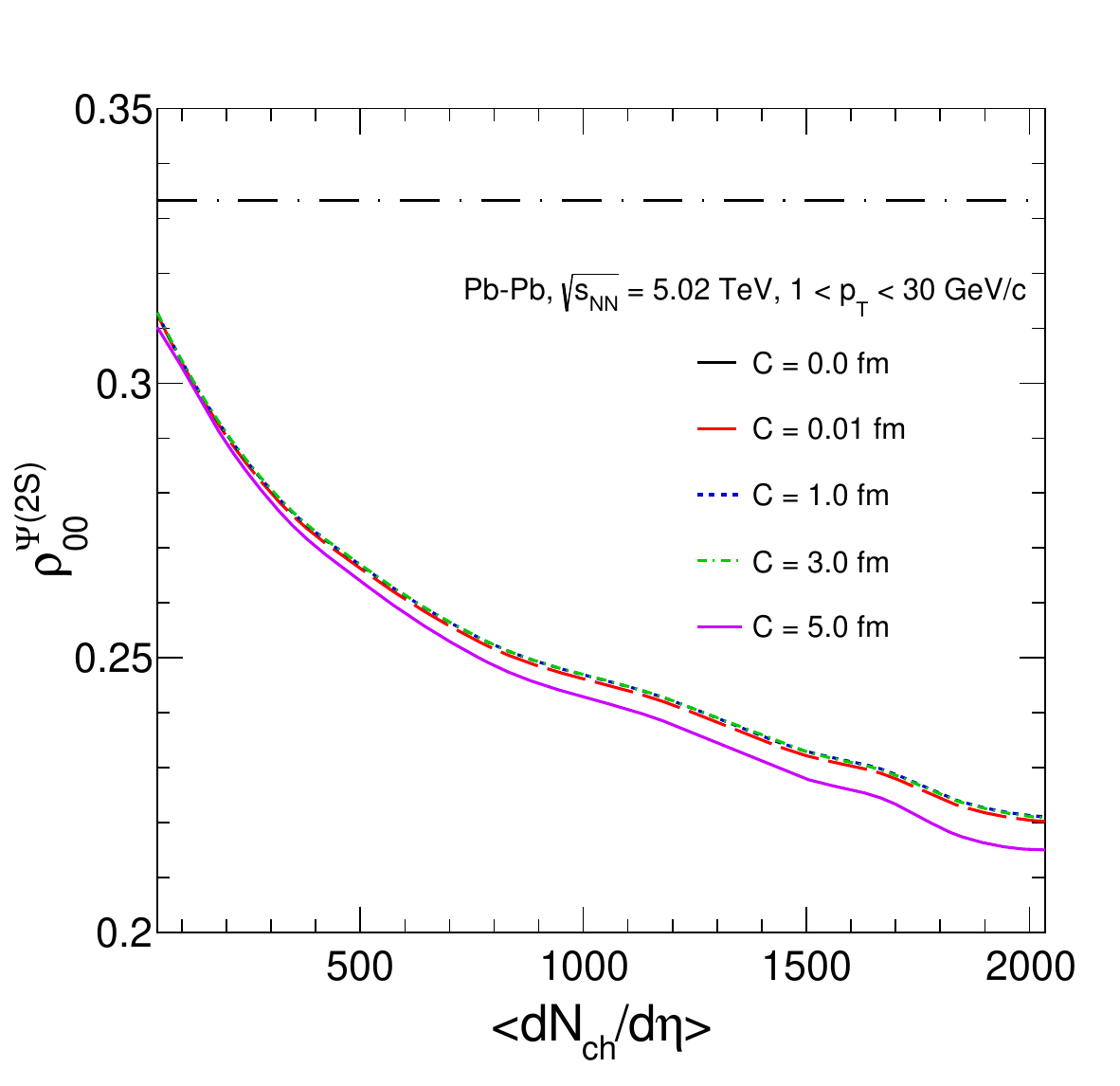}
\includegraphics[scale = 0.4]{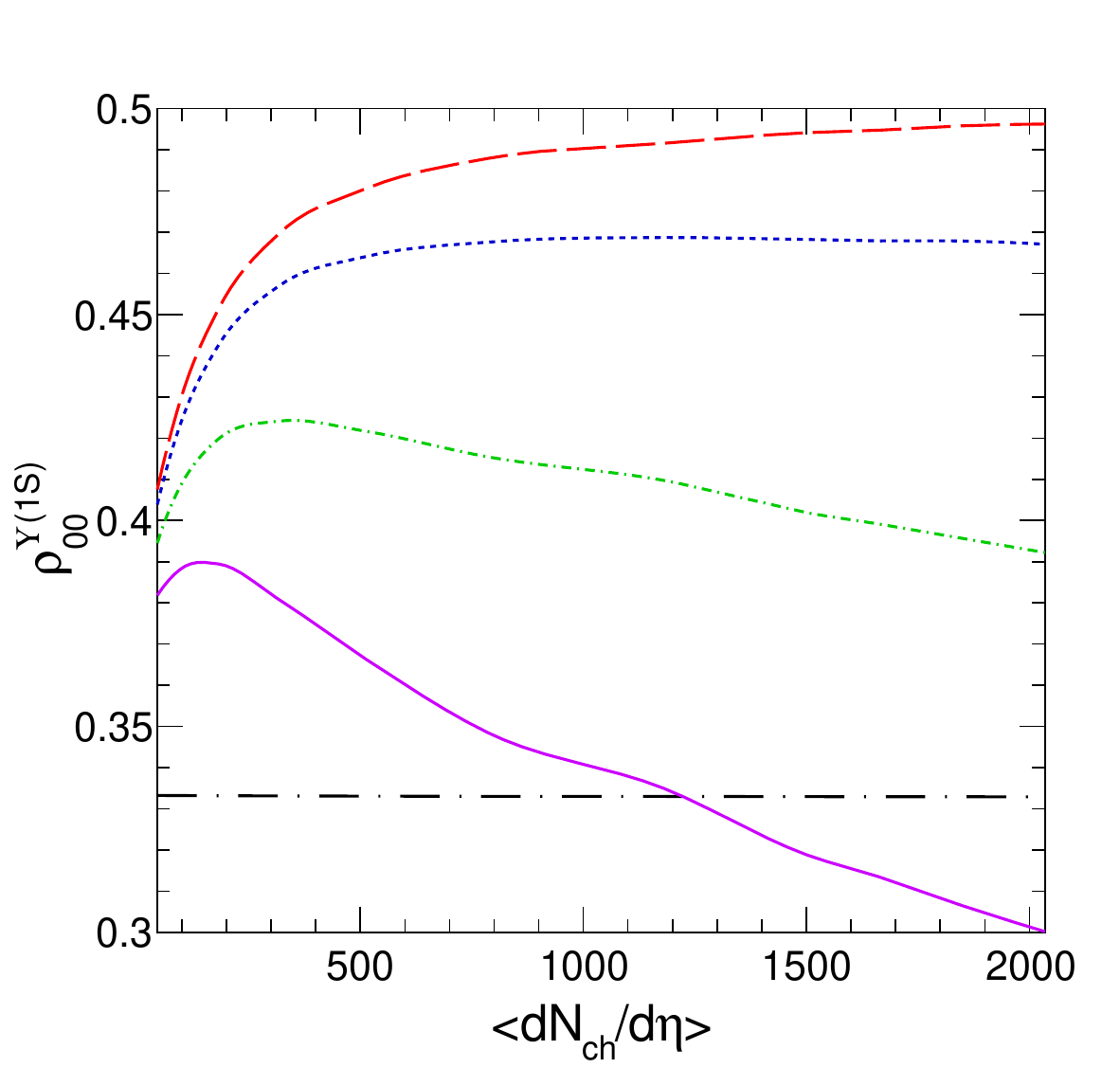}
\includegraphics[scale = 0.4]{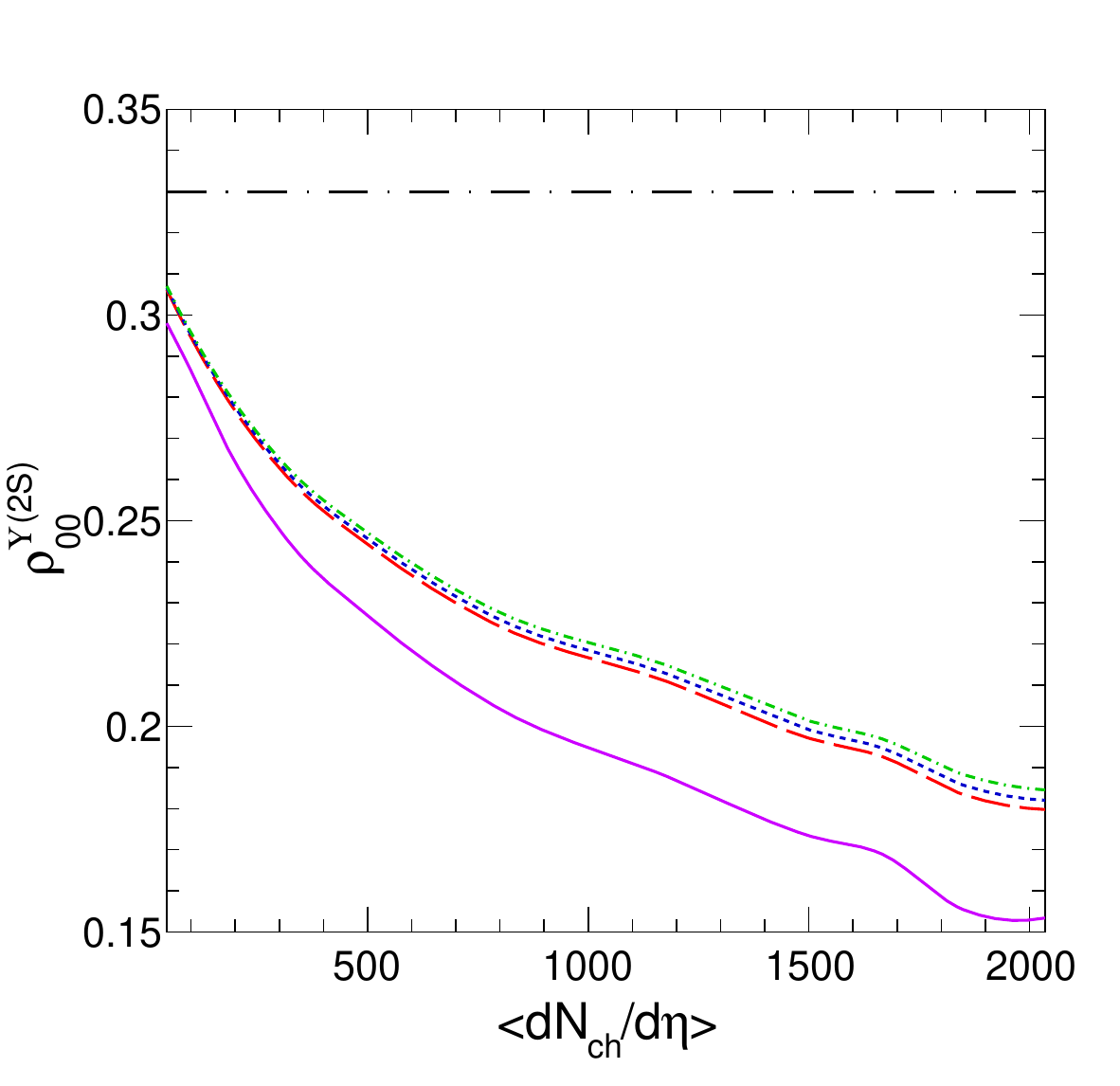}
\caption{(Color online) The spin alignment observable $\rho_{00}$ as a function of charged particle
multiplicity ($\langle dN_{\rm ch}/d\eta \rangle$) for $J/\psi$ (upper left), $\psi$(2S) (upper right),
$\Upsilon$(1S) (lower left), and $\Upsilon$(2S) (lower right) in Pb--Pb collisions at $\sqrt{s_{\rm
NN}}$ = 5.02 TeV integrated in the transverse momentum range 1$ < p_{\rm T} <$ 30 GeV/c for
different values of the circulation parameter, $C$.}
\label{fig:RhovsMult}
\end{figure*}

The decay width illustrated in Fig.~\ref{fig:GammaDwoC} shows the  $p_{\rm T}$ dependence for
$J/\psi$ (upper left), $\psi$(2S) (upper right), $\Upsilon$(1S) (lower left), and $\Upsilon$(2S)
(lower right) in Pb–Pb collisions at $\sqrt{s_{\rm NN}} = 5.02$ TeV in the absence in-medium
rotation ($C = 0$ fm) at $T$ = 175, 300, and 400 MeV. The results indicate a strong sensitivity of
the dissociation rate to both temperature and momentum. At $T$ = 400 MeV, the $J/\psi$ exhibits an
enhanced dissociation rate, which gradually decreases with increasing $p_{\rm T}$. The sudden drop
in $\Gamma_{\rm D}$ around $p_{\rm T} \sim $ 10 GeV/c reflects that $T_{\rm eff} \ll T = 400$ MeV.
Further, at $T$ = 300, and 175 MeV, the effect of $T_{\rm eff}$ significantly diminishes and
$\Gamma_{\rm D}$ decreases as a function of $p_{\rm T}$. A similar trend is evident for other
quarkonium states, including $\psi$(2S), $\Upsilon$(1S), and $\Upsilon$(2S). In particular, at
temperatures of 300 and 175 MeV, $\psi$(2S) exhibits a higher dissociation probability than $J/\psi$
across the entire $p_{\rm T}$ range. This can be attributed to the comparatively lower binding
energy of $\psi$(2S) relative to $J/\psi$. Consequently, the reduced survival probability of
$\psi$(2S) further supports the sequential melting scenario in the  QGP medium. Furthermore, due to
the higher mass of $\Upsilon$(1S) and $\Upsilon$(2S) states, which experience relatively stable
effective temperatures as they traverse through the medium, as illustrated in Fig.~\ref{fig:Teff}.
As a result, the dissociation probability  for $\Upsilon$(1S) exhibits a modest decreasing trend
towards higher $p_{\rm T}$. In contrast, for $\Upsilon$(2S), a slight increase in dissociation
probability is observed with rising $p_{\rm T}$.  Overall $p_{\rm T}$ dependence of $\Gamma_{\rm D}$
reflects the combined effects of the quarkonium decay width and its effective in-medium temperature.
These contrasting behaviors offer important insights into the interplay between quarkonium binding
energy and medium effects in heavy-ion collisions. \\

Further, Fig.~\ref{fig:GammaDwCchar} and Fig.~\ref{fig:GammaDwCbott} demonstrate the impact of
medium rotation on the  $\Gamma_{\rm D}$, as a function of $p_{\rm T}$ for charmonium  and
bottomonium states, respectively. In the presence of a vortical background, the degeneracy in a
quarkonium state is caused by spin-vorticity coupling. This implies that for $m_{j} = \pm 1$ state,
the change in net $\Gamma_{\rm D}$ is prominent, which is an artifact of the term
($\sim m_{j}C/r^{2}$) appearing in the effective Hamiltonian. The spin–vorticity coupling alters the
eigen energies and radial wave functions in a manner that depends on the sign of $m_{j}$. The
spin-vorticity coupling term ($\sim m_{j}C/r^{2}$) is expected to alter the short-distance ($
r\rightarrow$ 0 ) behavior of quarkonium wave functions. At a large distance, the behavior of
quarkonium wave functions is negligible. For the QCD vacuum potential with $m_{j}=+1$, the
additional term due to spin-orbit coupling lowers the effective potential barrier, thereby
increasing the spatial extent of the wave function, or making it more peaked at short distances.
Consequently, the dissociation probability rises, leading to a noticeably larger total decay width
$\Gamma_{\rm D}$ compared to the non-rotating case. In contrast, for $m_{j} = -1$, the rotational
term carries the opposite sign, effectively making the potential slightly more confining. This
suppresses the sensitivity of the quarkonium wave function at short distances. For the $m_{j}=0$
state,  the quarkonium wave function is not affected due to the direct coupling to the rotation
term, and naturally follows a trend similar to hydrogen  atom wave functions. However, with the
medium-modified color-singlet potential for quarkonia, the nature of the quarkonium wave function
strongly depends on temperature and effective coupling constant.  Further, the dissociation of the
quarkonium state in medium is led by the imaginary part of  the potential, which induces the Landau
damping. Therefore, the net decay width of a quarkonium state is a complex interplay between the
temperature-dependent quarkonium wave function and its imaginary part of the complex potential.  It
is observed that the behavior of the quarkonium wave function at $m_{j}=-1$ and $m_{j}=0$ is almost
similar. As a result, the net decay widths for the $m_{j}=0$ and $m_{j}=-1$ states remain closely
aligned for both charmonium and bottomonium.\\

Meanwhile, the splitting between the $m_{j}= +1$ and the other two spin projections for $\Gamma_{\rm
D}$ reflects the intrinsic asymmetry introduced by the $m_{j}C$ term in a rotating medium. This
rotational effect is more pronounced for charmonium than bottomonium because the latter has a
stronger binding, which reduces the relative impact of rotational distortions of the wave function.
Nonetheless, for most cases, medium rotation preferentially enhances the dissociation of the
$m_{j}=+1$ state in comparison with $m_{j}=0$ and $m_{j}=-1$ states. Furthermore, the change in the
$\Gamma_{\rm D}$ pattern with $p_{\rm T}$ in Fig.~\ref{fig:GammaDwCchar} and
Fig.~\ref{fig:GammaDwCbott} is arises due to the additional contribution of $T_{\rm eff}$ as
function of $p_{\rm T}$, which follows the similar explanation as Fig.~\ref{fig:GammaDwoC}.\\

\begin{figure*}[htbp]
\centering
\includegraphics[scale = 0.4]{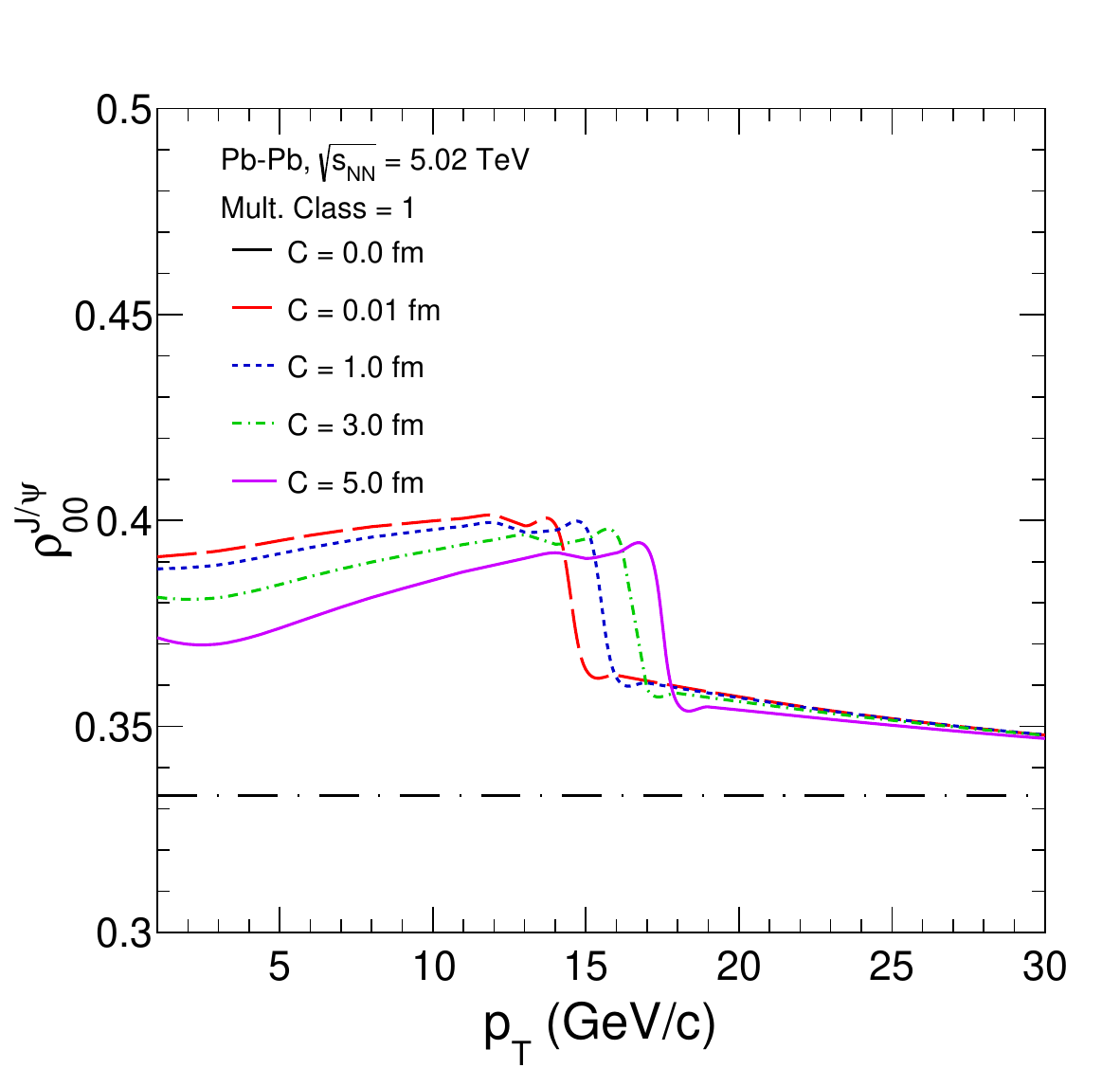}
\includegraphics[scale = 0.4]{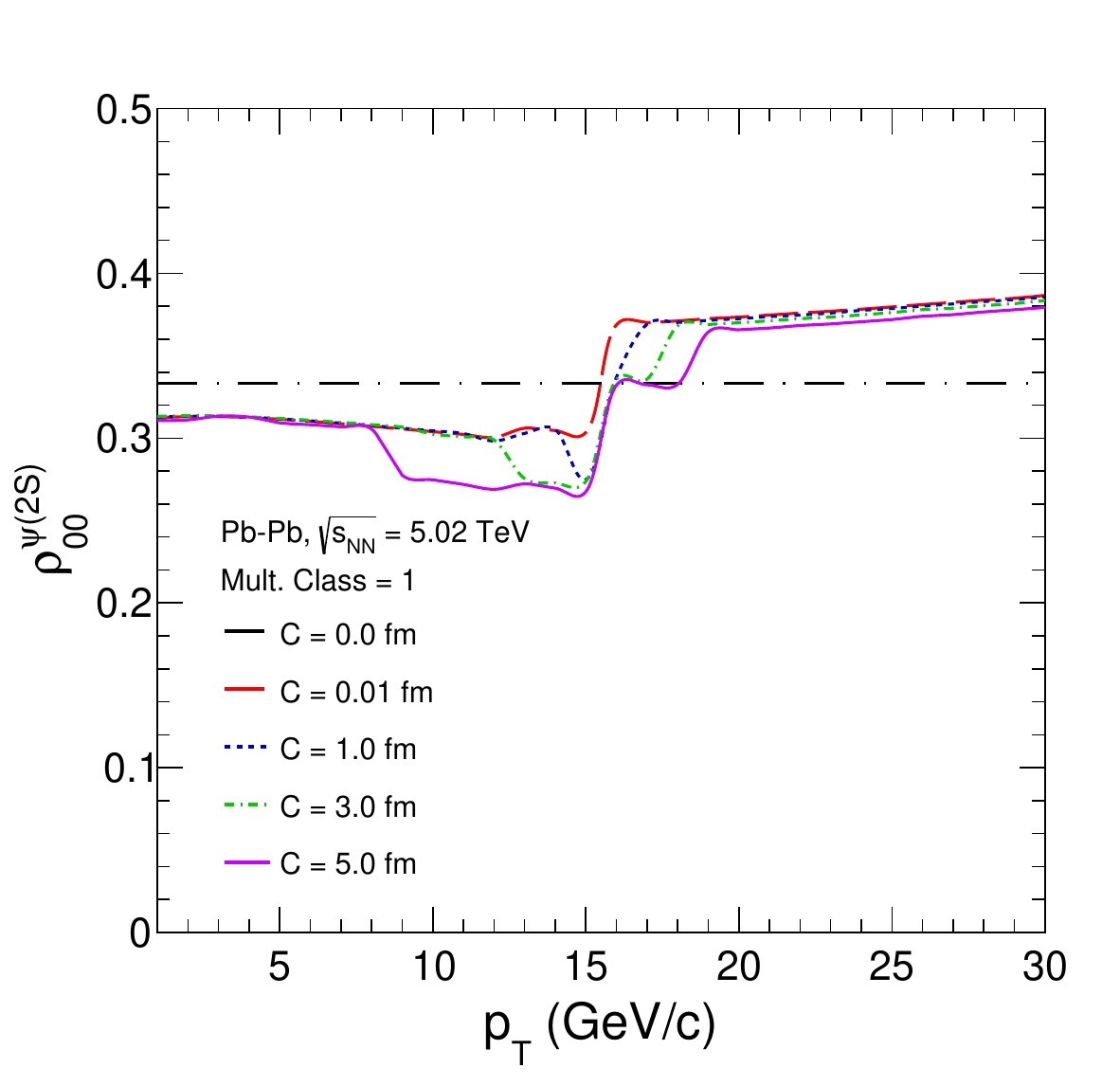}
\includegraphics[scale = 0.4]{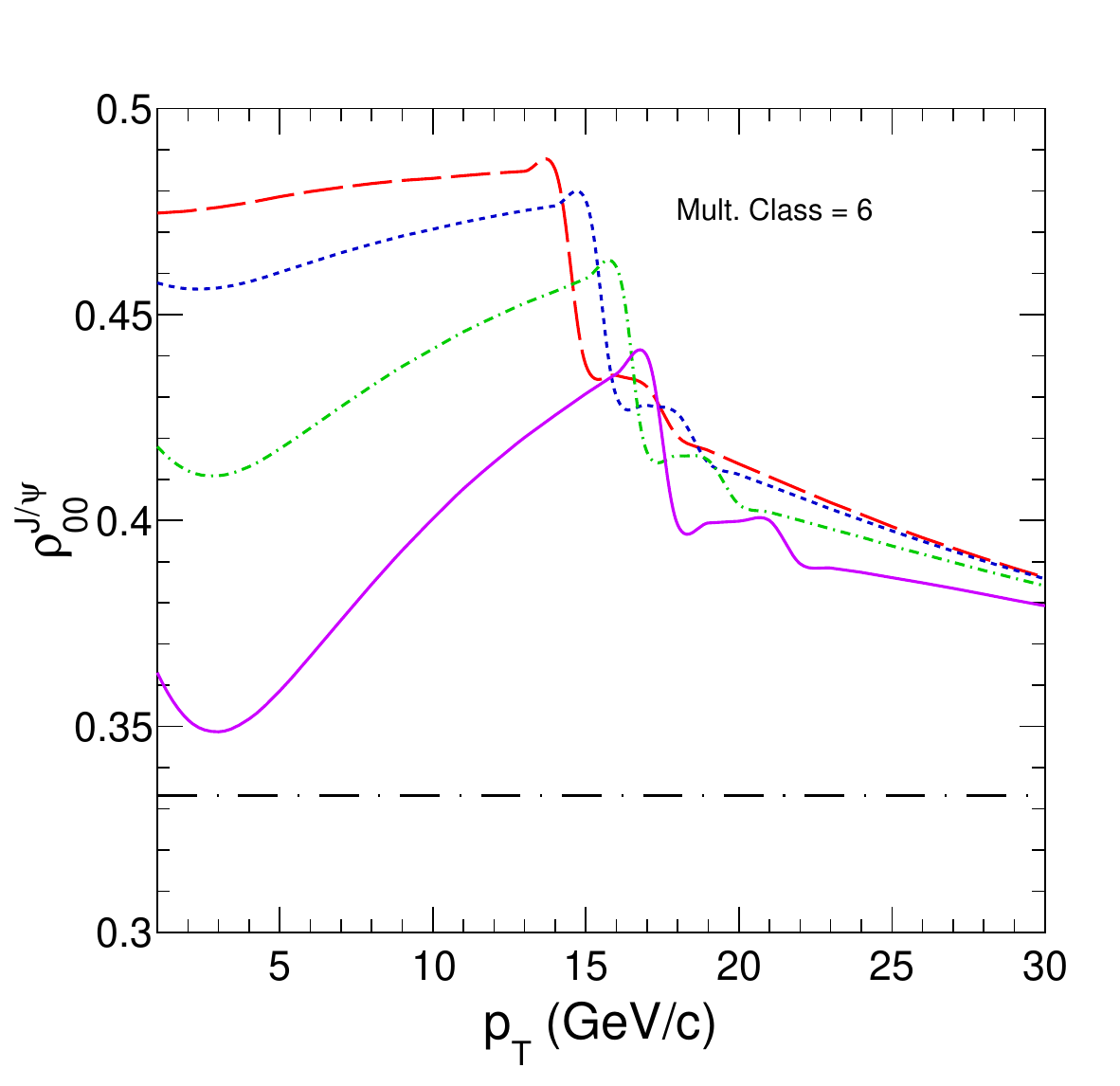}
\includegraphics[scale = 0.4]{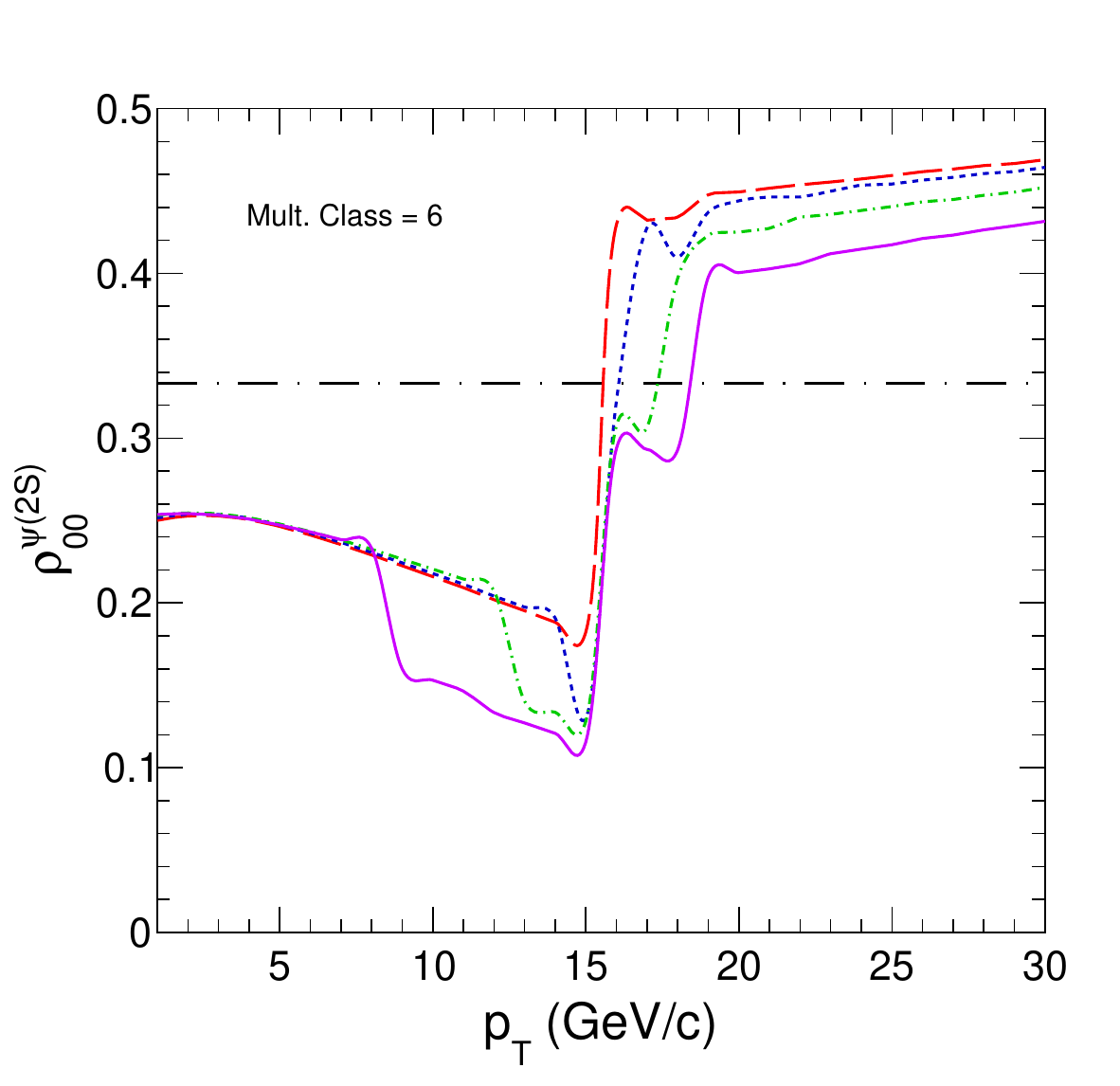}
\includegraphics[scale = 0.4]{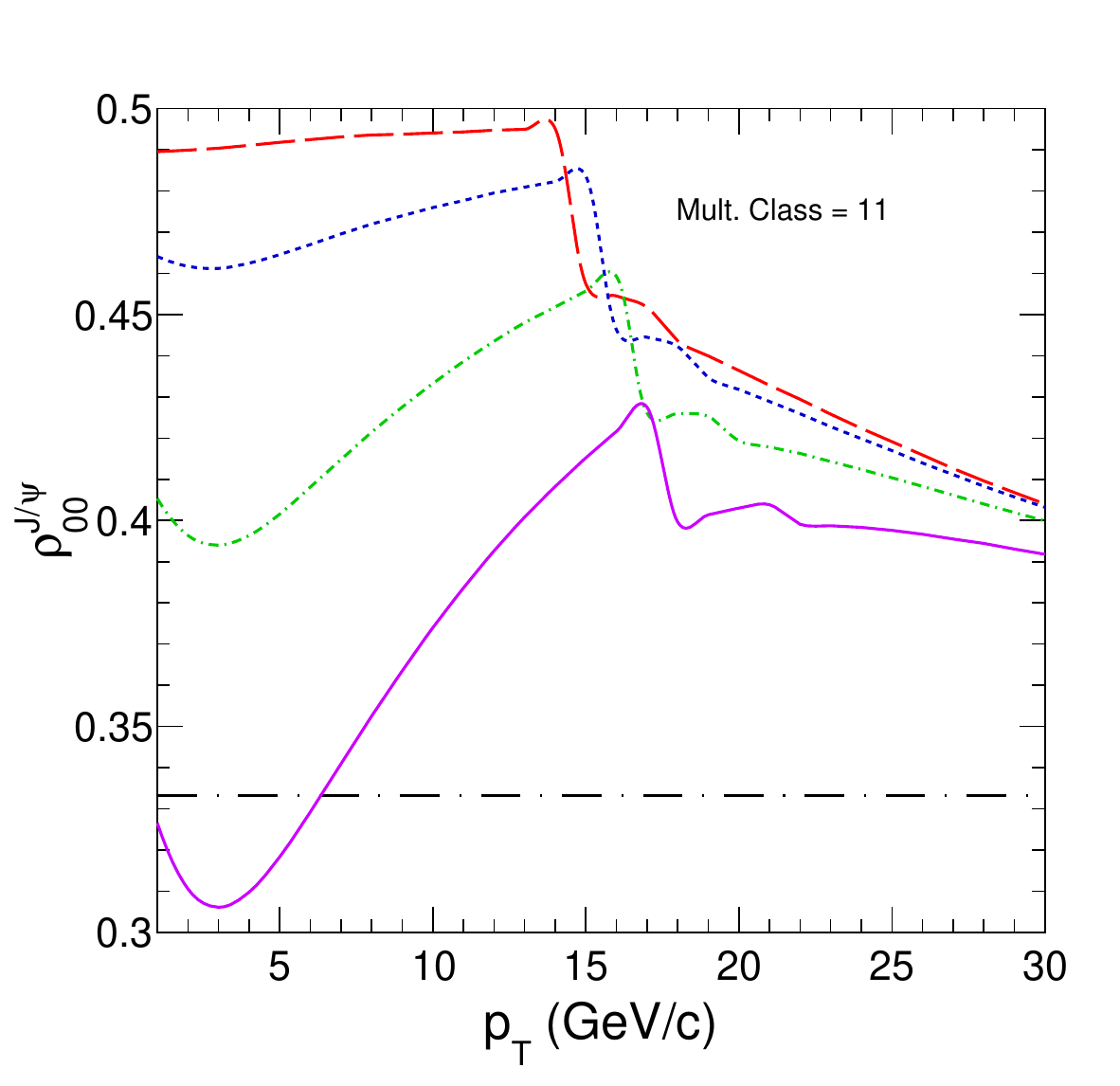}
\includegraphics[scale = 0.4]{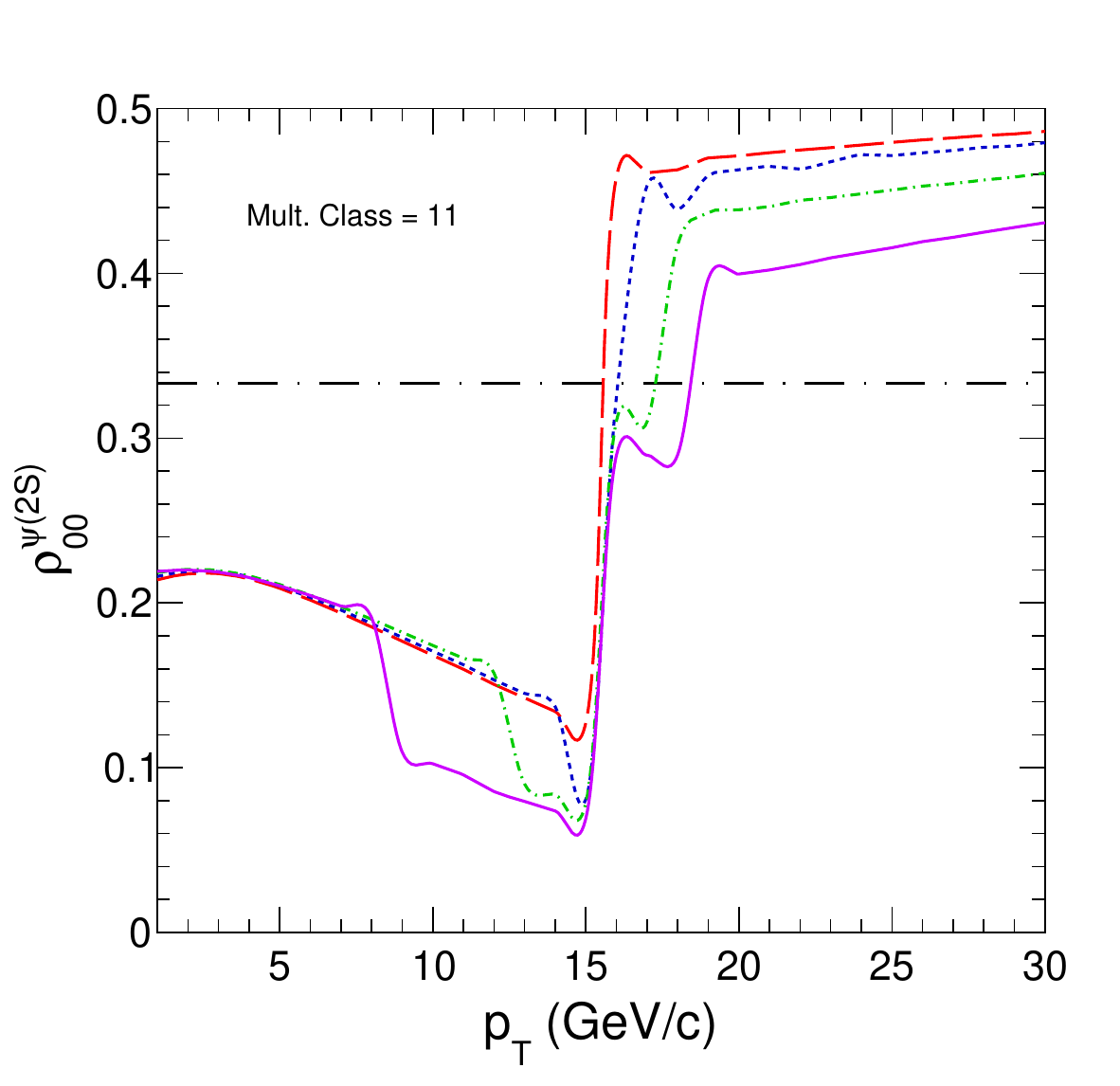}
\caption{(Color online) The spin alignment observable $\rho_{00}$ as a function of transverse
momentum  ($p_{\rm T}$) for $J/\psi$ (left panel) and $\psi$(2S) (right panel) in Pb--Pb collisions
at $\sqrt{s_{\rm NN}}$ = 5.02 TeV for various values of the circulation parameter, $C$ in three
multiplicity classes.}
\label{fig:CharRhovspT}
\end{figure*}

Figure~\ref{fig:RhovsMult} illustrates $00^{\rm th}-$elements of the spin density matrix,
$\rho_{00}$, as a function
of charged particle multiplicity ($\langle dN_{\rm ch}/d\eta \rangle$) for $J/\psi$ (upper left), $\psi$(2S)
(upper right),
$\Upsilon$(1S) (lower left), and $\Upsilon$(2S) (lower right) in Pb--Pb collisions at $\sqrt{s_{\rm
NN}}$ = 5.02
TeV at mid-rapidity. We have obtained the charged particle multiplicity dependent $\rho_{00}$ for quarkonium state 
by averaging over $p_{\rm T}$ in the  transverse momentum $p_{\rm T}$ range 1 $ \leq p_{\rm T} \leq $ 30 GeV/c with the 
$p_{\rm T}$-distribution function $1/E_{\rm T}^{4}$~\cite{Singh:2021evv}. The $p_{\rm T}$-integrated charged particle multiplicity dependent spin alignment is given as,

\begin{equation}
\rho_{00} (b) = \frac{\bigintssss_{p_{\rm Tmin}}^{p_{ \rm Tmax}} dp_{\rm T} \; \rho_{00}(p_{\rm T},
b)/(p_{\rm T}^{2} + M_{nl}^{2})^{2}}{\bigintssss_{p_{\rm Tmin}}^{p_{\rm Tmax}} d p_{\rm T}/(p_{\rm
T}^{2} + M_{nl}^{2})^{2}}
\end{equation}

The results corresponding to observable $\rho_{00}$ are presented for various values of the circulation parameter $C$, 
which represents the strength of the rotation of the medium. In the absence of the experimental constraint on the $C$, we 
chose a set of $C$ values spreading from 0.01 fm to 5.0 fm. These chosen values represent the theoretical uncertainty for 
the prediction of $\rho_{00}$. The dashed horizontal line at $\rho_{00}=1/3$ corresponds to an unpolarized scenario and 
serves as the baseline for comparison. In the $C=0$ case, corresponding to an irrotational medium,
all quarkonium states
yield $\rho_{00}=1/3$ independent of charged particle multiplicity, as expected in the absence of spin-vorticity coupling. 
Finite values of $C$ induce a clear deviation from this baseline, indicating the  medium rotation
leads to spin-dependent
dissociation and consequently generates spin alignment. Now at non-zero $C$, the $\rho_{00}$ for 1S
states like  $J/\psi$ and $\Upsilon$(1S),  remains above $1/3$ over the entire multiplicity range
for small and intermediate values of $C$, implying a relatively less dissociation of $m_j=0$ spin
state. At low multiplicities, where the medium temperature is relatively smaller, the effect of
vorticity is more pronounced and therefore, $\rho_{00}$ increases with $\langle dN_{\rm ch}/d\eta
\rangle$. As the multiplicity increases, the rise in-medium temperature enhances the relative
dissociation corresponding to $m_j=\pm 1$ degenerate states. While the effect of spin--vorticity
coupling induced due to small values of C, remains marginal and becomes almost ineffective at
higher temperatures or higher multiplicities. The combined dynamics of dissociation of $m_j=0,
\text{and}  \pm 1$ degenerate states, along with spin--vorticity coupling respective to small $C$
values, leads to a saturation $\rho_{00}$ at high multiplicities. For larger values of $C$, the
enhanced splitting among the spin-dependent decay widths increases the dissociation of the $m_j=0$
state, causing $\rho_{00}$ to approach the unpolarized limit for $J/\psi$ and a change in spin
alignment orientation for $\Upsilon$(1S) at high multiplicities.\\

Further, Fig.~\ref{fig:RhovsMult} depicts that quarkonium 2S states like $\psi$(2S) and
$\Upsilon$(2S) exhibit $\rho_{00}<1/3$ across the full multiplicity range for all nonzero values of
$C$. This behavior reflects the comparable small binding of the $2S$ states, which makes them
more susceptible to the thermal effects of the medium rather than rotational effects. As a
consequence, the medium temperature dominates over rotational effects except for the substantially
large values of C, and leads to enhanced dissociation of the $m_j=0$ component relative to the
$m_j=\pm1$ states. The $\rho_{00}$ for quarkonium $2S$ states shows a marginal dependence of $C$
over the whole range of multiplicity.
Consequently, $\rho_{00}$ decreases monotonically for $\psi$(2S) and $\Upsilon$(2S) with increasing charged particle multiplicity, indicating a transverse spin alignment. Conclusively,  Fig.~\ref{fig:RhovsMult} demonstrates that the multiplicity dependence of $\rho_{00}$ arises from the complex dynamics between medium vorticity and temperature-driven dissociation. While rotation-induced spin splitting plays a significant role for tightly bound $1S$ quarkonium states, thermal effects dominate for the loosely bound $2S$ states. This state-dependent behavior highlights the sensitivity of quarkonium spin alignment to both the binding energy of the quarkonia and the vortical structure of the QGP.\\

\begin{figure*}[htbp]
\centering
\includegraphics[scale = 0.4]{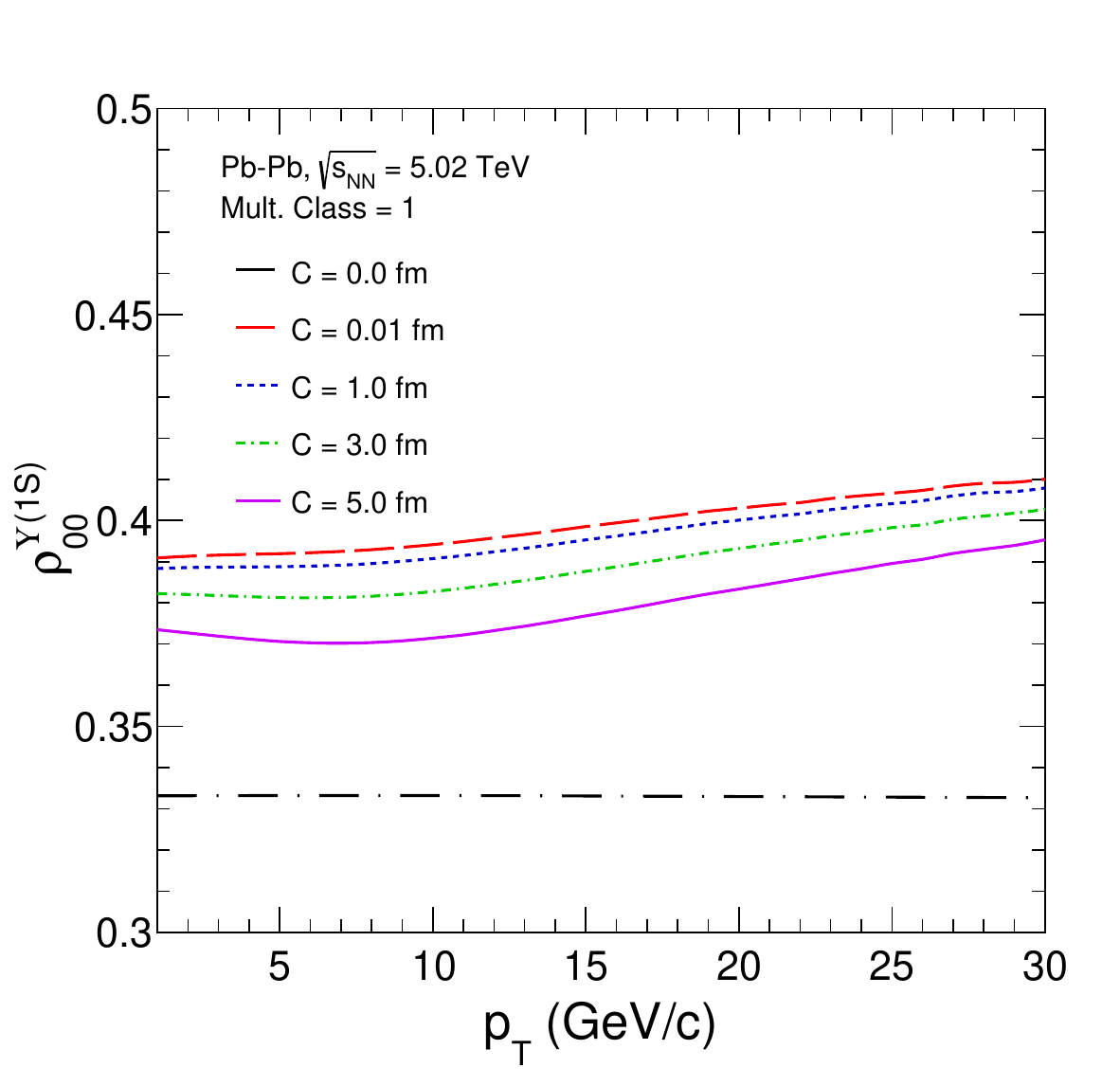}
\includegraphics[scale = 0.4]{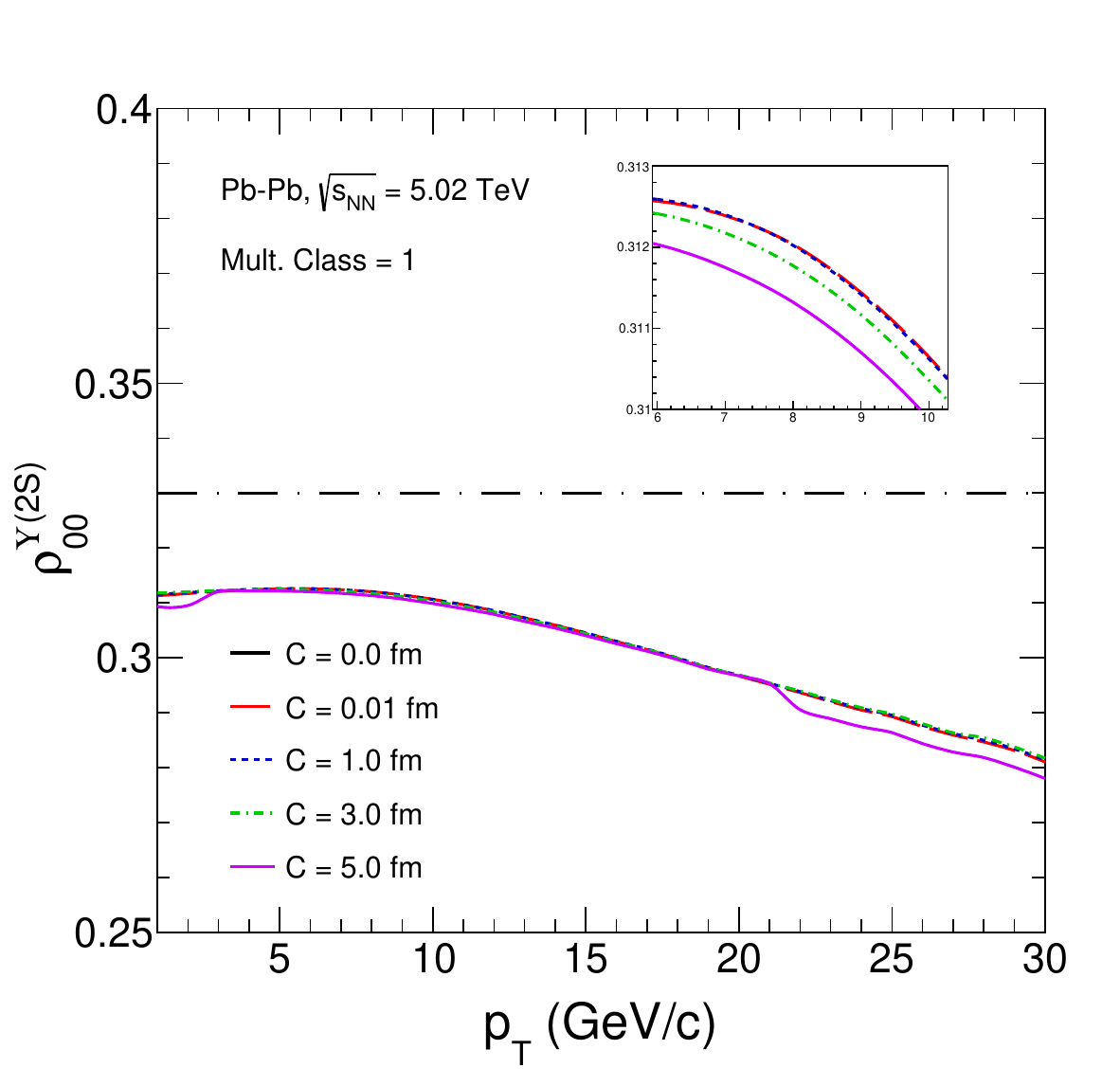}
\includegraphics[scale = 0.4]{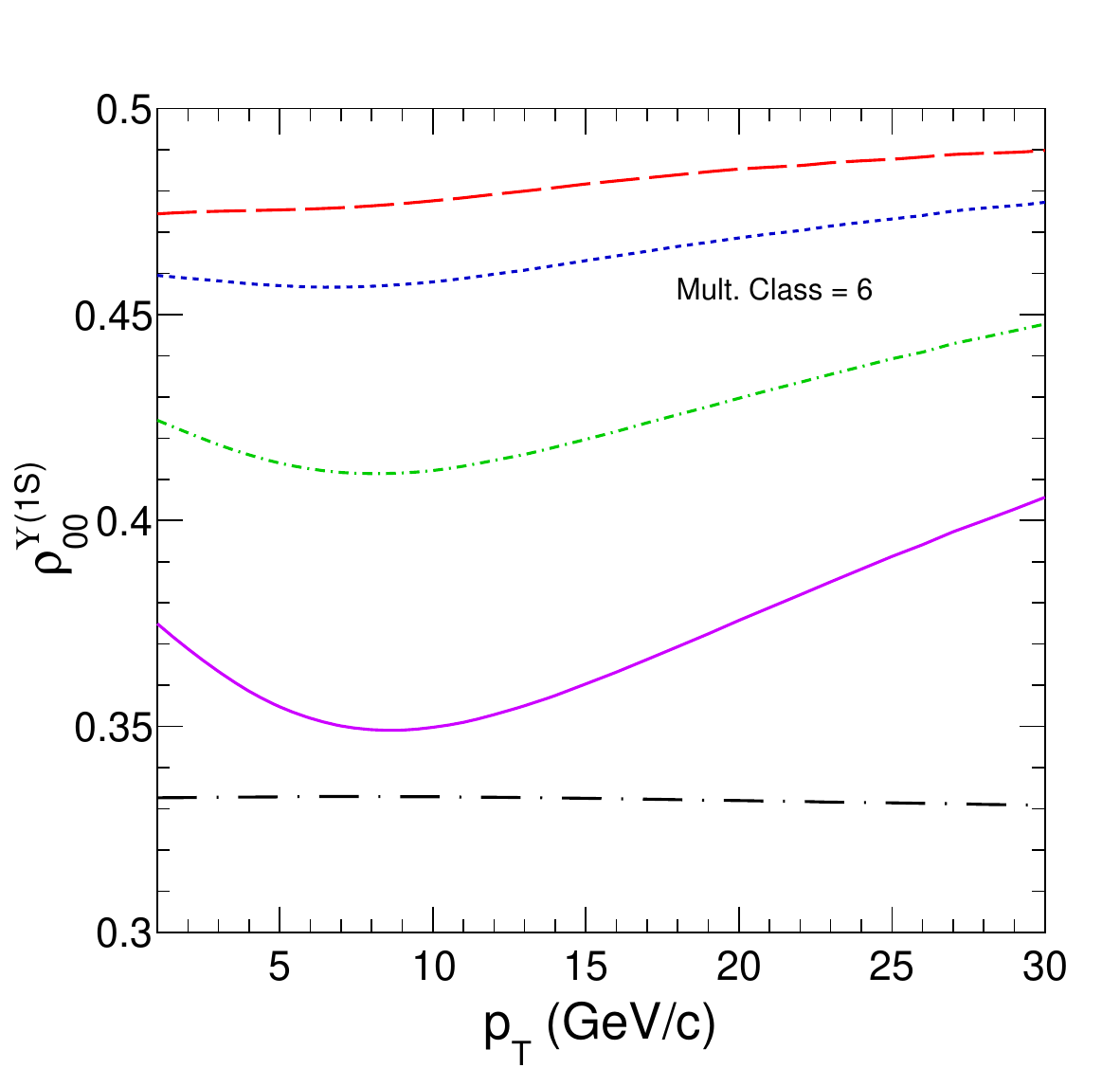}
\includegraphics[scale = 0.4]{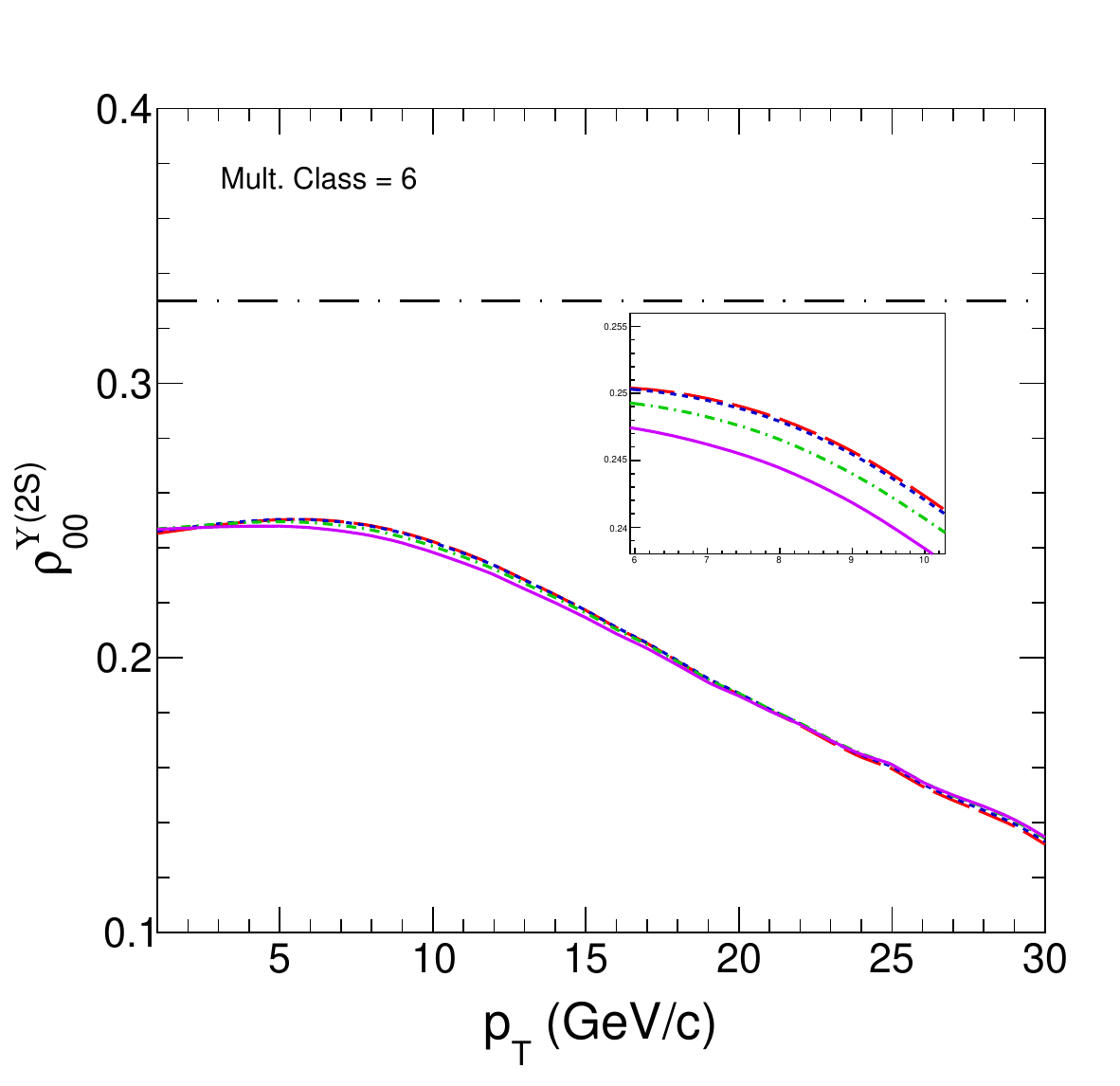}
\includegraphics[scale = 0.4]{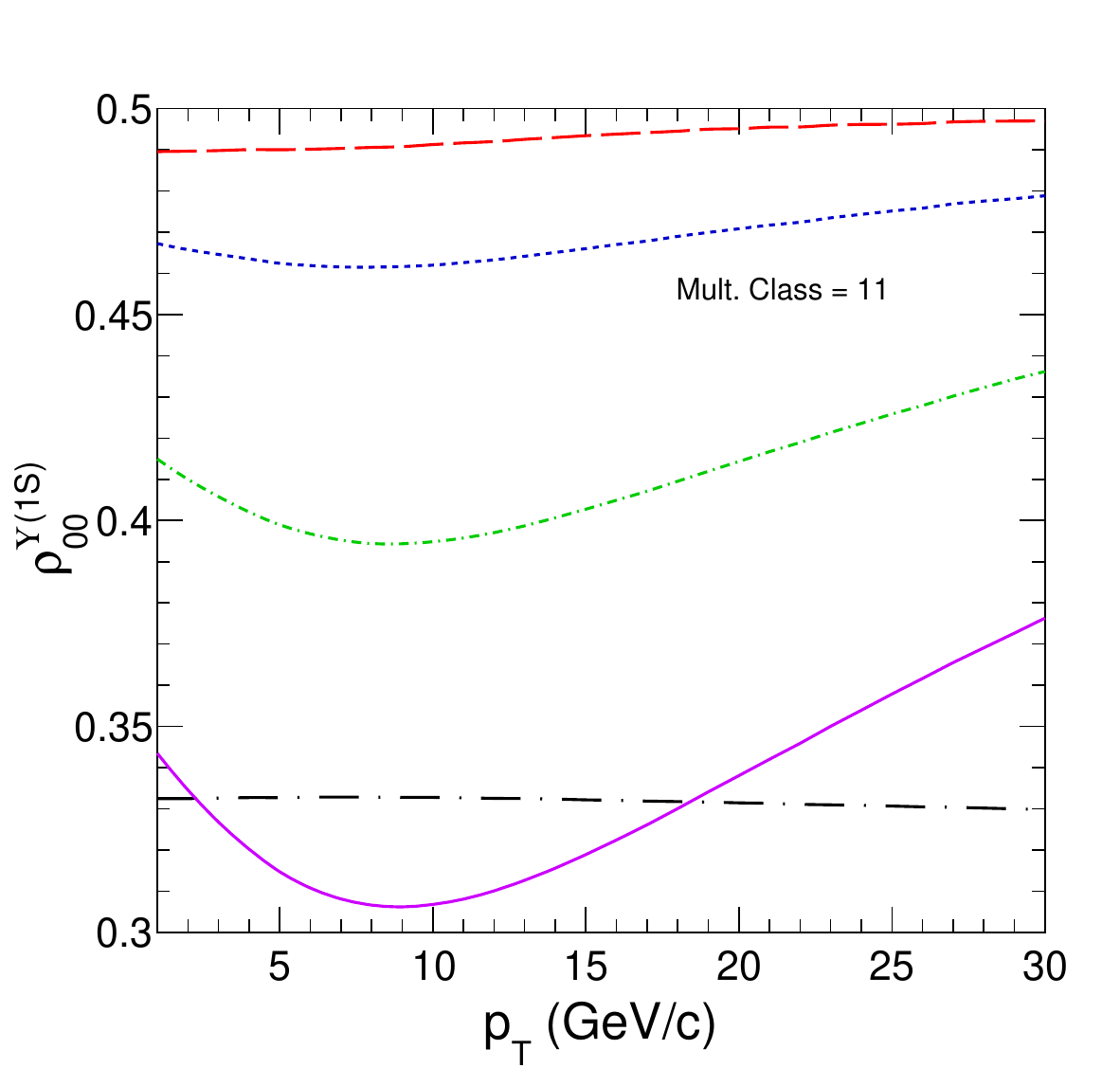}
\includegraphics[scale = 0.4]{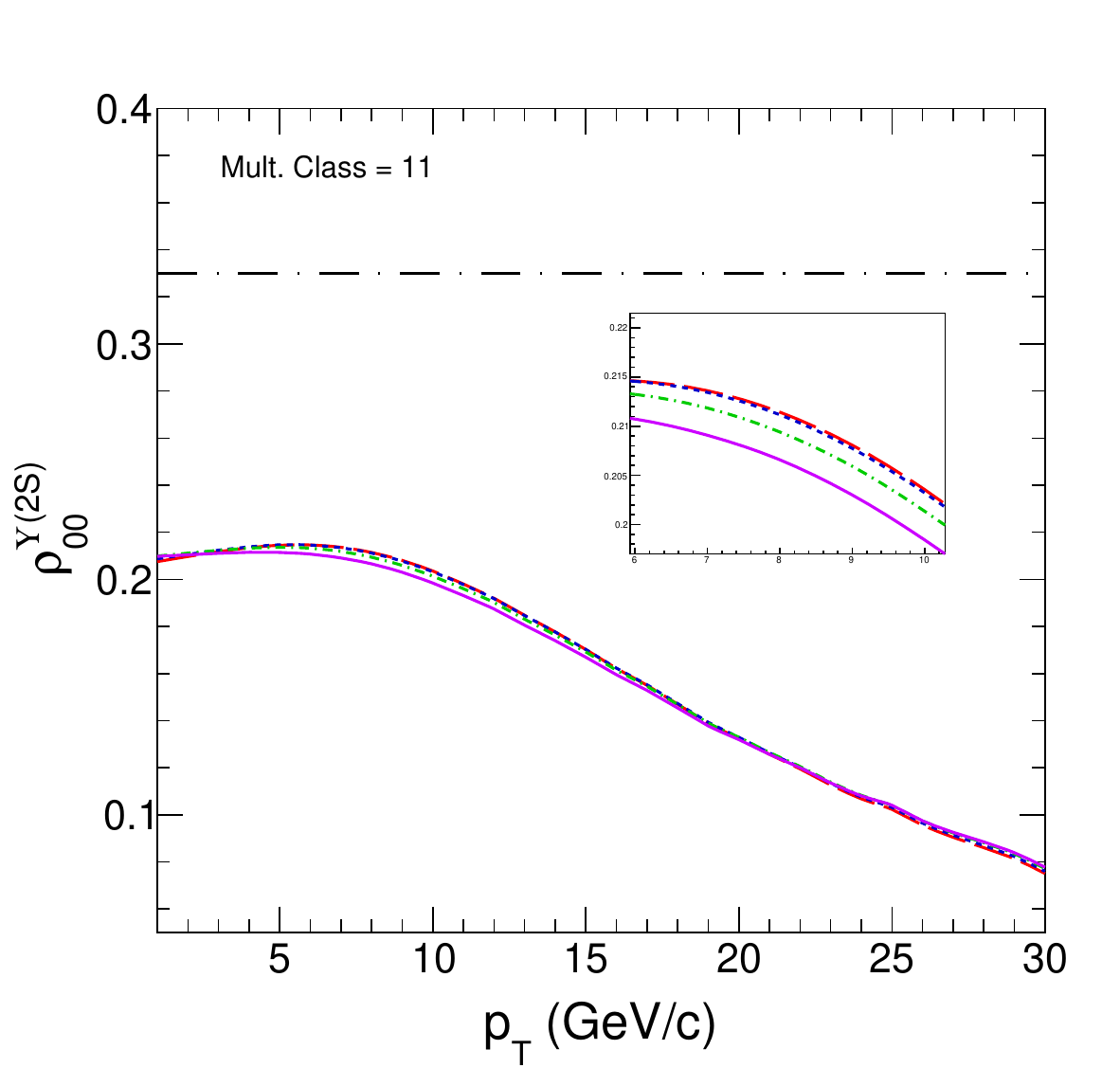}
\caption{(Color online) The spin alignment observable $\rho_{00}$ as a function of transverse momentum  ($p_{\rm 
T}$) for $\Upsilon$(1S) (left panel) and $\Upsilon$(2S) (right panel) in Pb--Pb collisions at $\sqrt{s_{\rm NN}}$ = 
5.02 TeV for various values of the circulation parameter, $C$, in three multiplicity classes.}
\label{fig:BottomRhovspT}
\end{figure*}

Figure ~\ref{fig:CharRhovspT} depicts the $p_{\rm T}$-dependent  $\rho_{00}$ for $J/\psi$ (left
panels) and $\psi$(2S) (right panels) for three multiplicities classes or centralities in Pb$-$Pb
collisions at $\sqrt{s_{\rm NN}}=5.02$ TeV. For the lowest multiplicity class (upper panels),
$\rho_{00}$ exhibits a non-monotonic dependence on $p_{\rm T}$. At low transverse momentum,
$\rho_{00}$ remains above the unpolarized limit, exhibiting a reduced dissociation of the $m_j=0$
state. With increasing $p_{\rm T}$, $\rho_{00}$ increases and reaches a maximum around $p_{\rm
T}\simeq 15$~GeV/$c$, followed by a gradual decrease toward $\rho_{00}=1/3$ at higher $p_{\rm T}$.
This behavior directly mirrors the $p_{\rm T}$ dependence of the net decay width $\Gamma_{\rm D}$,
which is governed by the interplay between collisional damping, gluonic dissociation, and the
momentum-dependent effective temperature. For higher multiplicity classes (middle and lower
panels), the $p_{\rm T}$ dependence of $\rho_{00}$ develops a characteristic structure at low
transverse momentum. In this region, $\rho_{00}$ initially decreases with increasing $p_{\rm T}$ up
to $p_{\rm T}\sim 3$~GeV/$c$, followed by a rising trend at intermediate and high $p_{\rm T}$. This
feature is an artifact of the $T_{\rm eff}$. Since $T_{\rm eff}$ enters explicitly into the
imaginary part of the in-medium potential, this non-monotonic behavior is directly transmitted to
$\Gamma_{\rm D}$ and consequently to $\rho_{00}$.\\

For a fixed multiplicity class, increasing the circulation parameter $C$ systematically reduces
$\rho_{00}$ across the entire $p_{\rm T}$ range  without modifying the overall shape of the
distribution. This reflects the fact that spin--vorticity coupling alters the relative dissociation
probabilities of the spin-projected states through the
$m_j C$ term in the effective Hamiltonian, while the dominant $p_{\rm T}$ dependence continues to be driven by $T_{\rm eff}$ and $\Gamma_{\rm D}$. The influence of vorticity remains modest for $C\lesssim 1$~fm and becomes more pronounced only at larger values of $C$.\\

The $\psi$(2S) results shown in the right panels of Fig.~\ref{fig:CharRhovspT} demonstrate that for
all multiplicity classes, $\rho_{00}^{\psi(2S)}$ remains below the limit $\rho_{00}=1/3$ over the
entire $p_{\rm T}$ range, implying a preferential dissociation of the $m_j=0$ spin state. The
$p_{\rm T}$ dependence of $\rho_{00}$ exhibits a characteristic non-monotonic structure,
particularly at higher multiplicities. At low transverse  momentum, $\rho_{00}$ decreases with
increasing $p_{\rm T}$ up to $p_{\rm T}\sim 10$--15~GeV/$c$, followed by an increase at  $p_{\rm T}
\geq 15$ GeV/c.  This behavior reflects the comparatively weaker binding of the $\psi$(2S) state
and sensitivity to the $T_{\rm eff}$, which makes it  more susceptible to medium-induced
dissociation effects. As a result, temperature-driven dissociation for $\psi$(2S) dominates over
rotational effects, leading to  a reduced survival probability of the $m_j=0$ component relative to
the $m_j=\pm 1$ states. It consequently shapes the spin alignment of $\psi$(2S) over low to high
multiplicities. For a fixed multiplicity class, the dependence of $\rho_{00}$  for ${\psi(2S)}$ on
the circulation parameter $C$ remains weak across the entire $p_{\rm T}$ range. It reflects that
spin-vorticity coupling for $\psi$(2S) provides only a subleading correction to the dominant thermal
dissociation mechanism. Even at larger values of $C$, the modification of the spin-dependent decay
widths is insufficient to overcome the strong temperature sensitivity associated with the small
binding energy of the $\psi$(2S) state.\\

In Fig.~\ref{fig:BottomRhovspT} spin alignment of $\Upsilon$(1S) and $\Upsilon$(2S) as a function of
$p_{\rm T}$ are shown for chosen $C$ values and multiplicity classes in Pb$-$Pb collisions at
$\sqrt{s_{\rm NN}}=5.02$ TeV.  At low multiplicity, $\rho_{00}$ for $\Upsilon$(1S) increases with
$p_{\rm T}$, consistent with a reduction of the net decay width $\Gamma_{\rm D}$ as the effective
temperature $T_{\rm eff}$ decreases at higher transverse momentum. At high multiplicities,
$\rho_{00}$ decreases up to $p_{\rm T} \leq 10$ GeV, followed by a gradual increase toward $p_{\rm
T} > 10$ GeV. It reflects the non-monotonic $p_{\rm T}$ dependence of $T_{\rm eff}$, which enhances
dissociation at low transverse momentum and reduces at higher $p_{\rm T}$. The systematic
enhancement of $\rho_{00}$ with increasing $C$ for all multiplicity classes indicates that
spin--vorticity coupling remains effective for the tightly bound $\Upsilon$(1S) state. In contrast,
the $\Upsilon$(2S) state exhibits $\rho_{00}<1/3$ across the whole $p_{\rm T}$-range, and
demonstrates the marginal dependence on both $p_{\rm T}$ and $C$. The smaller binding energy of the
$\Upsilon$(2S) state enhances its sensitivity to thermal dissociation, which dominates over
rotational effects and leads to a preferential dissociation of the $m_j=0$ component. Consequently,
variations in $T_{\rm eff}$ primarily govern the observed $p_{\rm T}$ dependence for $\Upsilon$(2S),
while spin-vorticity coupling introduces only marginal modifications.\\

Meanwhile, Fig.~\ref{fig:BottomRhovspT} demonstrates that the 
$p_{\rm T}$-dependent evolution of spin alignment for bottomonium states governed by the counterintuitive balance between temperature-driven dissociation and rotational effects. While the strongly bound $\Upsilon$(1S) state retains sensitivity to medium vorticity, the more weakly bound $\Upsilon$(2S) state is dominated by thermal dynamics, resulting in a qualitatively different and less vorticity-sensitive spin alignment pattern.

\section{Summary}
\label{sum}

In this work, we have systematically investigated the spin alignment of charmonium ($J/\psi$,
$\psi$(2S)) and bottomonium ($\Upsilon$(1S), $\Upsilon$(2S)) states in Pb--Pb collisions at
$\sqrt{s_{\rm NN}}=5.02$~TeV, focusing on dissociation-driven mechanisms in a rotating quark--gluon
plasma medium. By incorporating spin--vorticity coupling within an effective Hamiltonian framework
and employing a medium-modified complex potential, we evaluated spin-dependent decay widths arising
from collisional damping and gluonic dissociation. The space-time evolution of the medium is modeled
using second-order relativistic viscous hydrodynamics, and the temperature cooling profile and its
impact  on quarkonium dynamics are investigated.\\

Our results demonstrate that medium vorticity induces a spin-dependent modification of the
quarkonium decay widths through the $m_j C$ term in the effective Hamiltonian, which directly
translates into a measurable deviation of the spin  density matrix element $\rho_{00}$ from the
unpolarized baseline. For tightly bound $1S$ states, $J/\psi$ and $\Upsilon$(1S), $\rho_{00}$
predominantly exceeds $1/3$, indicating a relative stability of the $m_j=0$ spin projection and a
clear sensitivity to rotational effects, particularly at low to intermediate charged particle
multiplicities. In contrast, the excited $2S$ states, $\psi$(2S) and $\Upsilon$(2S), exhibit
$\rho_{00}<1/3$ across most of the explored kinematic range, reflecting their enhanced
susceptibility to temperature-driven dissociation and a reduced sensitivity to medium vorticity.
Further, $p_{\rm T}$-dependence of $\rho_{00}$ reveals a nontrivial structure governed by the
momentum-dependent effective temperature $T_{\rm eff}$, which enters explicitly into the imaginary
part of the in-medium potential. The resulting non-monotonic behavior of the net decay width
$\Gamma_{\rm D}$ with $p_{\rm T}$ is directly imprinted onto the spin alignment observable for both
charmonium and bottomonium states. A consistent hierarchy emerges across all observables, wherein
the interplay between binding energy, effective temperature, and rotational dynamics determines the
final spin alignment pattern.\\

Conclusively, this study establishes spin-dependent dissociation in a vortical QGP as a viable microscopic mechanism contributing to quarkonium spin alignment. The results provide a unified interpretation of the observed state, momentum, and multiplicity-dependent behavior of $\rho_{00}$, offering a complementary perspective to spin alignment mechanisms based solely on spin transport or hadronization effects. These findings strengthen the role of quarkonium spin observables as sensitive probes of both the thermal and vortical properties of deconfined QCD matter. \\

\subsection*{Future Outlook}

The extension of the present work can further advance the understanding of quarkonium spin dynamics in relativistic 
heavy-ion collisions.

\begin{itemize}

\item A most general improvement involves embedding the current framework into a 3+1D 
viscous hydrodynamic, which would allow a more realistic treatment of spatial vorticity fluctuations and local temperature 
gradients. Such an approach would enable a quantitative assessment of the sensitivity of
$\rho_{00}$ to the vorticity.

\item In addition, extending the present analysis to small collision systems, such as high-multiplicity $p$-Pb and 
$p-p$ collisions, would be particularly valuable in light of recent experimental indications of collective behavior. A 
comparative study across system sizes could disentangle genuine medium-induced effects from initial-state contributions.

\item From an experimental perspective, differential measurements of quarkonium spin alignment in multiple polarization 
frames and with finer binning in transverse momentum and centrality will be crucial for constraining theoretical models. 
The  present results provide concrete predictions for the state-dependent sensitivity of $\rho_{00}$ to medium rotation, 
which can be tested in future high-statistics datasets from the LHC and RHIC facilities.

\item Finally, incorporating additional microscopic effects, such as regeneration, feed-down contributions, and possible 
spin-dependent recombination mechanisms, would further allow a more comprehensive description of quarkonium spin alignment 
across the evolution of the medium. Together, these developments will help establish quarkonium spin observables as 
precision tools for probing the vortical structure and microscopic dynamics of the
quark--gluon plasma.
\end{itemize}

\section*{Acknowledgement}
Bhagyarathi Sahoo acknowledges the financial aid from CSIR, Government of India. The authors gratefully acknowledge the DAE-DST, Government of India, funding under the mega-science project "Indian Participation in the ALICE experiment at CERN" bearing Project No. SR/MF/PS-02/2021-IITI (E-37123).


\begin{thebibliography}{}

\bibitem{Matsui:1986dk}
T.~Matsui and H.~Satz,
Phys. Lett. B \textbf{178}, 416 (1986).


\bibitem{ALICE:2012jsl}
B.~Abelev \textit{et al.} [ALICE Collaboration],
Phys. Rev. Lett. \textbf{109}, 072301 (2012).

\bibitem{CMS:2012bms}
S.~Chatrchyan \textit{et al.} [CMS Collaboration],
JHEP \textbf{05}, 063 (2012).

\bibitem{ALICE:2013xmt}
B.~B.~Abelev \textit{et al.} [ALICE Collaboration],
Phys. Lett. B \textbf{728}, 216 (2014).
[erratum: Phys. Lett. B \textbf{734}, 409 (2014).]

\bibitem{ALICE:2016ccg}
J.~Adam \textit{et al.} [ALICE Collaboration],
Phys. Rev. Lett. \textbf{116}, 132302 (2016).

\bibitem{ALICE:2016fzo}
J.~Adam \textit{et al.} [ALICE Collaboration],
Nature Phys. \textbf{13}, 535 (2017).

\bibitem{CMS:2016fnw}
V.~Khachatryan \textit{et al.} [CMS Collaboration],
Phys. Lett. B \textbf{765}, 193 (2017).

\bibitem{ALICE:2024vzv}
S.~Acharya \textit{et al.} [ALICE Collaboration], 
[arXiv:2411.09323].


\bibitem{STAR:2017ckg}
L.~Adamczyk \textit{et al.} [STAR Collaboration],
Nature \textbf{548} 62 (2017).

\bibitem{STAR:2020xbm}
J.~Adam \textit{et al.} [STAR Collaboration],
Phys. Rev. Lett. \textbf{126}, 162301 (2021).

\bibitem{STAR:2023nvo}
M.~I.~Abdulhamid \textit{et al.} [STAR Collaboration],
Phys. Rev. C \textbf{108}, 014910 (2023).

\bibitem{ALICE:2019aid}
S.~Acharya \textit{et al.} [ALICE Collaboration],
Phys. Rev. Lett. \textbf{125}, 012301 (2020).

\bibitem{STAR:2022fan}
M.~S.~Abdallah \textit{et al.} [STAR Collaboration],
Nature \textbf{614}, 244 (2023).

\bibitem{ALICE:2025cdf}
S.~Acharya \textit{et al.} [ALICE Collaboration],
JHEP \textbf{10}, 094 (2025).

\bibitem{Liang:2004ph}
Z.~T.~Liang and X.~N.~Wang,
Phys. Rev. Lett. \textbf{94}, 102301 (2005).
[erratum: Phys. Rev. Lett. \textbf{96} (2006), 039901]

\bibitem{Liang:2004xn}
Z.~T.~Liang and X.~N.~Wang,
Phys. Lett. B \textbf{629}, 20 (2005).

\bibitem{Karpenko:2016jyx}
I.~Karpenko and F.~Becattini,
Eur. Phys. J. C \textbf{77}, 213 (2017).

\bibitem{Alzhrani:2022dpi}
S.~Alzhrani, S.~Ryu and C.~Shen,
Phys. Rev. C \textbf{106}, 014905 (2022).

\bibitem{Teryaev:2017wlm}
O.~V.~Teryaev and V.~I.~Zakharov,
Phys. Rev. D \textbf{96}, 096023 (2017).

\bibitem{Sheng:2019kmk}
X.~L.~Sheng, L.~Oliva and Q.~Wang,
Phys. Rev. D \textbf{101}, 096005 (2020).
[erratum: Phys. Rev. D \textbf{105}, 099903 (2022).]

\bibitem{Sheng:2022wsy}
X.~L.~Sheng, L.~Oliva, Z.~T.~Liang, Q.~Wang and X.~N.~Wang,
Phys. Rev. Lett. \textbf{131}, 042304 (2023).


\bibitem{Sahoo:2024yud}
B.~Sahoo, C.~R.~Singh and R.~Sahoo,
Eur. Phys. J. C \textbf{85}, 580 (2025).

\bibitem{Sahoo:2024egx}
B.~Sahoo, C.~R.~Singh and R.~Sahoo,
Phys. Scripta \textbf{100}, 065310 (2025).

\bibitem{Sahoo:2025bkx}
B.~Sahoo, C.~R.~Singh and R.~Sahoo,
[arXiv:2506.09405].

\bibitem{Sahoo:2023oid}
B.~Sahoo, D.~Sahu, S.~Deb, C.~R.~Singh and R.~Sahoo,
Phys. Rev. C \textbf{109}, 034910 (2024).

\bibitem{LHCb:2025rxf}
R.~Aaij \textit{et al.} [LHCb],
Phys. Rev. D \textbf{112}, 112022 (2025).

\bibitem{STAR:2025jwc}
 The STAR Collaboration,
[arXiv:2509.17487].

\bibitem{ALICE:2020iev}
S.~Acharya \textit{et al.} [ALICE Collaboration],
Phys. Lett. B \textbf{815}, 136146 (2021).

\bibitem{ALICE:2022dyy}
S.~Acharya \textit{et al.} [ALICE Collaboration],
Phys. Rev. Lett. \textbf{131}, 042303 (2023).

\bibitem{Becattini:2007sr}
F.~Becattini, F.~Piccinini and J.~Rizzo,
Phys. Rev. C \textbf{77}, 024906 (2008).

\bibitem{Bhalerao:2020ulk}
R.~S.~Bhalerao,
Eur. Phys. J. ST \textbf{230}, 635 (2021).

\bibitem{Singh:2018wdt}
C.~R.~Singh, S.~Ganesh and M.~Mishra,
Eur. Phys. J. C \textbf{79}, 147 (2019).

\bibitem{Singh:2021evv}
C.~R.~Singh, S.~Deb, R.~Sahoo and J.~e.~Alam,
Eur. Phys. J. C \textbf{82}, 542 (2022).

\bibitem{Singh:2025xrd}
C.~R.~Singh, P.~Bagchi, R.~Sahoo and J.~e.~Alam,
Phys. Rev. D \textbf{112}, 014017 (2025).


\bibitem{Singh:2015eta}
C.~R.~Singh, P.~K.~Srivastava, S.~Ganesh and M.~Mishra,
Phys. Rev. C \textbf{92}, 034916 (2015).

\bibitem{Ganesh:2016kug}
S.~Ganesh, C.~R.~Singh and M.~Mishra,
J. Phys. G \textbf{45}, 035003 (2018).

\bibitem{Hatwar:2020esf}
N.~Hatwar, C.~R.~Singh, S.~Ganesh and M.~Mishra,
Phys. Rev. C \textbf{104}, 034905 (2021).

\bibitem{Hwa:1985xg}
R.~C.~Hwa and K.~Kajantie,
Phys. Rev. D \textbf{32}, 1109 (1985).

\bibitem{ALICE:2015juo}
J.~Adam \textit{et al.} [ALICE Collaboration],
Phys. Rev. Lett. \textbf{116}, 222302 (2016).

\bibitem{Muronga:2003ta}
A.~Muronga,
Phys. Rev. C \textbf{69}, 034903 (2004).

\bibitem{gam} 
H. Grad, Commun. Pure Appl. Math. {\bf 2} , 331 (1949).

\bibitem{mis1} 
W. Israel, Ann. Phys. (N.Y.) {\bf 100}, 310 (1976).

\bibitem{mis2} 
J. M. Stewart, Proc. R. Soc. London A {\bf 357}, 59 (1977).

\bibitem{mis3} W. Israel and J. M. Stewart, Ann. Phys. (N.Y.) {\bf 118}, 341 (1979).

\bibitem{Anan} 
Anandan, J.,  Suzuki, J. (n.d.). Quantum Mechanics in a Rotating Frame. \url{https://doi.org/https://arxiv.org/pdf/quant-ph/0305081.pdf}

\bibitem{Gordan}
Gordan Baym, Lectures on Quantum Mechanics (1969). \url{https://archive.org/details/lecturesonquantu0000baym/page/n611/mode/2up}

\bibitem{Griffth}
David J. Griffths, Introduction to Quantum Mechanics.
\url{https://www.fisica.net/mecanica-quantica/Griffiths%20-%20Introduction%20to%20quantum%20mechanics.pdf}

\bibitem{nendzig}
F. Nendzig and G. Wolschin, Phys. Rev. C {\bf 87}, 024911 (2013).

\bibitem{Peskin:1979va}
M.E. Peskin, Nucl. Phys. B {\bf 156}, 365 (1979).

\bibitem{Bhanot:1979vb}
 G. Bhanot, M.E. Peskin, Nucl. Phys. B {\bf 156}, 391 (1979).
 
\bibitem{Grandchamp:2001pf}
L. Grandchamp, R. Rapp, Nucl. Phys. A {\bf 709}, 415 (2002).

\bibitem{Brambilla:2008cx}
N. Brambilla, J. Ghiglieri, A. Vairo, P. Petreczky, Phys. Rev. D {\bf 78},
014017 (2008).
\end{thebibliography}
\end{document}